\documentclass[12pt,a4paper]{JHEP3}

\usepackage{amssymb, amsmath, amsopn, amsthm, dsfont}
\usepackage{epsfig}

\allowdisplaybreaks

\newcommand{\cN}{\mathcal{N}}

\DeclareMathOperator{\Tr}{Tr}

\newcommand{\g}{\gamma}

\newcommand{\cM}{\mathcal{M}}

\newcommand{\cP}{\mathcal{P}}

\newcommand{\beq}{\begin{equation}}
\newcommand{\eeq}{\end{equation}}

\newcommand{\lp}{\left(}
\newcommand{\rp}{\right)}
\newcommand{\ls}{\left[}
\newcommand{\rs}{\right]}

\renewcommand{\d}{\delta}
\newcommand{\e}{\epsilon}

\newcommand{\m}{\mu}
\newcommand{\n}{\nu}

\newcommand{\om}{\omega}
\newcommand{\s}{\sigma}

\newcommand{\hph}[1]{{\hphantom{#1}}}

\newcommand{\U}{{\operatorname{U}}}
\newcommand{\hlf}{{\frac{1}{2}}}
\newcommand{\p}{\partial}
\newcommand{\non}{\nonumber}
\newcommand{\mcC}{{\mathcal{C}}}

\newcommand{\be}{\begin{equation}}
\newcommand{\ee}{\end{equation}}
\newcommand{\bea}{\begin{eqnarray}}
\newcommand{\eea}{\end{eqnarray}}
\newcommand{\bean}{\begin{eqnarray*}}
\newcommand{\eean}{\end{eqnarray*}}
\newcommand{\nn}{\nonumber}

\newcommand{\hx}{\hat{x}}
\newcommand{\hy}{\hat{y}}
\newcommand{\hatt}{\hat{t}}
\newcommand{\hr}{\hat{r}}
\newcommand{\hz}{\hat{z}}

\newcommand{\hX}{\hat{X}}
\newcommand{\hY}{\hat{Y}}
\newcommand{\hA}{\hat{A}}

\newcommand{\slashed}[1]{\displaystyle{\not} #1}

\newcommand{\ul}[1]{\underline{#1}}

\def\rmi{{\rm i}}

\newcommand{\ft}[2]{{\textstyle\frac{#1}{#2}}}
\newcommand{\vm}[1]{{\vec m}_{#1}}

\title{Higher Derivative Corrections and Central Charges from Wrapped M5-branes}

\author{Marco Baggio${}^{1}$, Nick Halmagyi${}^{2,3}$, Daniel R. Mayerson${}^4$, Daniel Robbins${}^4$, Brian Wecht${}^5$\\

{\tt baggiom@ethz.ch, halmagyi@lpthe.jussieu.fr, d.r.mayerson@uva.nl, d.g.robbins@uva.nl, b.wecht@qmul.ac.uk}\\
${}^1$ Institut f\"ur Theoretische Physik, ETH Zurich,
CH-8093 Z\"urich, Switzerland \\
${}^2$ Sorbonne Universit\'{e}s, UPMC Paris 06, UMR 7589, LPTHE, 75005, Paris, France \\
${}^3$ CNRS, UMR 7589, LPTHE, 75005, Paris, France \\
${}^4$ Institute for Theoretical Physics, University of Amsterdam,\\
Science Park 904, Postbus 94485, 1090 GL Amsterdam, The Netherlands \\
${}^5$ Centre for Research in String Theory, Queen Mary University of London, London E1 4NS, United Kingdom \\
}

\abstract{We compute four-derivative corrections to the AdS supergravity actions arising from the near-horizon geometry of $N$ M5-branes wrapped on either one or two Riemann surfaces. This setup features the novel presence of both gauged isometries as well as nontrivial hypermultiplets. We argue
that the 5d Chern-Simons terms receive not only higher-derivative corrections 
but also contributions from Killing vector parameters, which we find must also be corrected. We check the central charges found by our supergravity methods against the dual field theory results and find perfect agreement at leading and subleading order in $N$. Along the way, we find higher derivative corrections to general $AdS_5$ and $AdS_3\times \Sigma_g$ geometries.}

\preprint{}

\begin{document}
\allowdisplaybreaks
\section{Introduction}
\label{sec:intro}
Conformal anomalies provide an important tool for analyzing CFTs in various dimensions. In particular, they were instrumental in the early tests of the AdS/CFT correspondence, where the large-$N$ central charges of $\mathcal{N}=4$ SYM were successfully reproduced by a gravity computation in \cite{Henningson:1998gx}. Subsequently, the match was extended to finite $N$ in this theory and others by taking into account higher order effects in the bulk; theories with $1/N$ effects were discussed in \cite{Aharony:1999rz}, and theories with $1/N^2$ corrections were explored in \cite{Anselmi:1998zb,Bilal:1999ty,Bilal:1999ph,Mansfield:2000zw,Mansfield:2002pa,Mansfield:2003gs}. The higher-derivative corrections for related quiver theories were studied in \cite{Liu:2010gz,Ardehali:2013gra,Ardehali:2013xya,Ardehali:2013xla}.
In some cases, exact knowledge of the field theory central charges was used to infer the precise form of higher-derivative supergravity corrections, as in \cite{Tseytlin:2000sf}. 

In this paper, we are concerned with $\mathcal{N}=(0,2)$ and $\mathcal{N}=1$ superconformal field theories in two and four dimensions, respectively. In such theories, the stress-energy tensor is part of a supermultiplet that includes the $U(1)_R$ R-symmetry current. As a consequence, the central charges can be derived exactly from the knowledge of the $U(1)_R$ '��t Hooft anomalies. Since there might be many global $U(1)$ symmetries, the determination of the central charges is tantamount to identifying the correct $U(1)_R$ symmetry among the many possibilities. In \cite{Intriligator:2003jj} it was proven that the exact R-symmetry of $\mathcal{N}=1$ theories in four dimensions is the one that maximizes the central charge $a$. More recently, a similar result was proven for two-dimensional $\mathcal{N}=(0,2)$ theories \cite{Benini:2012cz}, where it was shown that the exact R-symmetry extremizes the right-moving central charge $c_R$. The supergravity duals of these procedures were  described first in \cite{Tachikawa:2005tq} and subsequently \cite{Hanaki:2006pj,Szepietowski:2012tb} for $a$-maximization and in \cite{Karndumri:2013iqa} for $c$-extremization at the two-derivative level.

Theories with these amounts of supersymmetry appear in numerous string theory constructions, where they describe the low-energy worldvolume physics of various brane configurations. Of particular interest to us will be setups consisting of $N$ parallel M5-branes, whose worldvolume is described by a six-dimensional $\mathcal{N}=(2,0)$ supersymmetric theory. Very little is known about this theory, but one  fruitful approach has been to wrap the branes around one or two Riemann surfaces and study the resulting low-energy effective theories \cite{Gaiotto:2009we,Gaiotto:2009hg,Alday:2009qq,Kanno:2009ga,Gaiotto:2010be,Tachikawa:2010vg,Nanopoulos:2010ga,Chacaltana:2010ks,Gaiotto:2011tf,Cecotti:2011rv,Tachikawa:2011yr,Alim:2011ae,Gaiotto:2011xs,Alim:2011kw}. This procedure can give rise to interesting four-dimensional $\mathcal{N}=1$ as well as two-dimensional $\mathcal{N}=(0,2)$ superconformal field theories. The exact central charges of these theories were computed by $a$-maximization \cite{Bah:2011vv,Bah:2012dg,Bah:2011je} and $c$-extremization \cite{Benini:2013cda,Benini:2012cz}.

The near-horizon limit of the backreacted geometries that arise from these M5-brane setups  interesting dual $AdS_5$ \cite{Maldacena:2000mw} and $AdS_3$ \cite{Gauntlett:2001jj} 
solutions in supergravity. The central charges can then be computed at leading order in $N$ using standard holographic techniques, and they agree with the exact CFT results. Typically, these leading order contributions scale as $N^3$. In this paper we wish to reproduce the first subleading (order $1/N^2$) corrections to these central charges by considering various higher-derivative corrections to the supergravity theory and computing the corrections to the $AdS_5$ and $AdS_3$ geometries describing the near-horizon limit of these brane setups. The case of M5-branes wrapped around 4-cycles of Calabi--Yau manifolds has been considered earlier in the literature \cite{Castro:2007sd}, where they give rise to $AdS_3 \times S^2$ solutions of ungauged 5d $\mathcal{N}=1$ supergravity. In our setups we will have to deal with two additional complications: the presence of gauged isometries and non-trivial hypermultiplets.

Since the eight-derivative corrections to 11d supergravity are not known in closed form, we will use a rather indirect strategy. We will focus on $\mathcal{N}=1$ five-dimensional supergravity, which is a consistent truncation of the eleven-dimensional theory that contains all the solutions of interest. The advantage of the five-dimensional formulation is the availability of powerful off-shell techniques which have made it possible to compute the supersymmetric completion of the $Weyl^2$ \cite{Hanaki:2006pj} as well as the $R^2$ \cite{Ozkan:2013nwa} terms.

Our strategy will be as follows: first we consider the well-known CP-odd eight-derivative correction of 11d supergravity $C_3 \wedge X_8$ to derive the subleading corrections to the five-dimensional Chern--Simons terms.  We then embed both the leading and subleading Chern--Simons terms in a fully supersymmetric five-dimensional Lagrangian. Finally, using off-shell techniques, we compute the corrections to the $AdS_5$ and $AdS_3 \times \Sigma_g$ geometries and reproduce the central charges of the dual theories.

A tantalizing outcome of our analysis is that on top of the ``explicit" higher-derivative corrections to the action, it is necessary to introduce $1/N$ corrections to the Killing vectors that gauge the global symmetries of the hypermultiplet sector. This can be seen as the gauged counterpart of $1/N$ corrections to the universal hypermultiplet geometry studied in \cite{Strominger:1997eb,Gunther:1998sc,Antoniadis:2003sw,RoblesLlana:2006ez,Alexandrov:2007ec} for compactifications of M-theory on Calabi--Yau manifolds.

The paper is organized as follows. In section \ref{sec:fieldthy} we review the main field theory results that we wish to reproduce from the supergravity side. In particular, we will briefly review the techniques of $a$-maximization and $c$-extremization and present the central charges for the IR SCFTs describing M5 branes wrapped on one or two Riemann surfaces. In section \ref{sec:sugrarev} we present the supergravity conventions and techniques that will be used throughout the paper. More importantly, we construct general $AdS_5$ and $AdS_3 \times \Sigma_g$ solutions in the presence of higher-derivative corrections. In section \ref{sec:hyperscalars} we specialize the considerations of the previous section to the case of M5-branes wrapped on one or two Riemann surfaces. This allows us to holographically reproduce the central charges of the dual SCFTs and derive the specific subleading corrections to the Killing vectors. We conclude in section 5 with some open problems and possible directions for future work.
\section{Field Theory}
\label{sec:fieldthy}

In this section, we review the results from two- and four-dimensional field theory that we will aim to reproduce from a supergravity perspective. 

\subsection{Four Dimensions}\label{sec:4DCFT}

The bosonic sector of the four-dimensional ${\cal N}=1$ superconformal algebra is $SO(4,2) \times U(1)_R$. The latter factor is the four-dimensional R-symmetry, the knowledge of which has many useful consequences. In particular, knowing the R-charges of a given theory allows the computation of the central charges $a$ and $c$, which are given by:
\begin{equation}
a = \frac{3}{32}\left [ 3{\rm Tr} R^3 - {\rm Tr} R \right ] \qquad c = \frac{1}{32}\left [ 9 {\rm Tr} R^3 - 5 {\rm Tr} R \right ].
\end{equation}
One advantage of these formulae is that the traces can sometimes be computed even in theories which have no known Lagrangian descriptions, such as theories that come from compactifying M5-branes. 

In many cases of practical interest, the R-symmetry is not immediately obvious. In particular, for any candidate symmetry $R_0$, it is possible that a putative superconformal R-symmetry could mix with non-R global symmetries $F_I$, yielding a family of R-symmetries $R_t (s^I) = R_0 + \sum_I s^I F_I$. To determine the values of $s^I$ that correspond to the unique superconformal R-symmetry, we must employ $a$-maximization \cite{Intriligator:2003jj}, and find the (local) maximum of $R_t(s^I)$. Since this cubic function can have at most one local maximum, this procedure uniquely determines the R-symmetry.

A useful alternate perspective on $a$-maximization can be found by using the anomaly polynomial. Recall that a chiral fermion in a four-dimensional theory with charge $q$ under a $U(1)$ global symmetry ${\cal F}$ has a six-form anomaly polynomial given by:
\begin{equation}
\label{sixform}
I_6 = {\rm ch}({\cal F}) \hat A(T) |_6 = \frac{q^3}{6}c_1({\cal F})^3 - \frac{q}{24}c_1({\cal F}) p_1(T),
\end{equation}
where $c_1({\cal F})$ is the first Chern class of ${\cal F}$ and $p_1(T)$ is the first Pontryagin class of the tangent bundle of the four-dimensional spacetime. For theories that come from M5-branes wrapped on a Riemann surface $\Sigma$, we have the advantage of knowing the 8-form anomaly polynomial $I_8$ of the $(2,0)$-theory, and can reproduce the four-dimensional anomaly polynomial by integrating $I_8$ over $\Sigma$. 
In particular, the M5-brane anomaly polynomial is given by \cite{Harvey:1998bx, Yi:2001bz, Intriligator:2000eq}: 
\begin{equation}
I_8 = \frac{r_G}{48} \left [ p_2(N) - p_2(T) + \frac14 \left ( p_1(T) - p_1(N) \right )^2 \right ] + \frac{r_G h_G (h_G+1)}{24}p_2(N),
\end{equation}
where $N$ and $T$ are the normal and tangent bundles,  $p_i$ is the $i^{\rm th}$ Pontryagin class, and $r_G$ and $h_G$ are the rank and Coexter number, respectively, of the type of $(2,0)$ theory being considered. For example, for the $A_{N-1}$ theory, $r_G = N-1$ and  $h_G = N$.  The dimension of the group is given by $d_G=r_G(h_G+1)$.

Consider wrapping an M5-brane on a Riemann surface $\Sigma$ whose normal bundle is $U(1)^2$, with Chern numbers $p$ and $q$.  Supersymmetry requires that $p+q=2\mathfrak{g}-2$, where $\mathfrak{g}$ is the genus of $\Sigma$.  It will be useful to parametrize the supersymmetric solutions via:
\be
p=(1+z)(\mathfrak{g}-1),\qquad q=(1-z)(\mathfrak{g}-1).
\ee
Some linear combination of these two $U(1)$'s is the R-symmetry, which can then potentially mix with a linear combination corresponding to a non-R symmetry. We can encode this mixing into an ambiguity in the individual Chern roots, which is then reflected in the coefficients of the 6-form anomaly polynomial we get when integrating $I_8$ over $\Sigma$ \cite{Bah:2012dg}. By identifying the coefficients of Eq.~(\ref{sixform}) with ${\rm Tr}\, R_t^3$ and ${\rm Tr} \,R_t$, we can then use $a$-maximization to find the superconformal R-symmetry. A major advantage of this technique is that it not only gives the exact ({\it i.e.}, not only large $N$) answer, but can also be used despite the absence of a four-dimensional Lagrangian.

The final result for the central charges is \cite{Bah:2012dg}:
\begin{align}
	a & = |\mathfrak{g}-1|r_G \frac{\zeta^3 + \kappa \s^3 - \kappa (1+\s)(9+21\s+9\s^2)z^2}{48(1+\s)^2 z^2},\\
	c & = |\mathfrak{g}-1|r_G \frac{\zeta^3 + \kappa \s^3 - \kappa (1+\s)(6 - \kappa \zeta + 17\s + 9\s^2)z^2}{48(1+\s)^2 z^2},
\end{align}
where the two parameters $\s$ and $\zeta$ are defined as:
\begin{align}
	\s &= h_G(1+h_G), & \zeta & = \sqrt{\s^2+(1+4\s + 3 \s^2)z^2},
\end{align}
while $\kappa = 1$ for $S^2$ and $\kappa = -1$ if the Riemann surface is hyperbolic ($\mathfrak{g}>1$). The case of $T^2$, with $\mathfrak{g}=1$ and $\kappa = 0$, must be treated separately and leads to the central charges:
\begin{align}
	a &= \frac{|z|}{48}\frac{r_G(1+3\s)^{3/2}}{\sqrt{1+\s}},\\
	c &= \frac{|z|}{48}\frac{r_G(2+3\s)\sqrt{1+3\s}}{\sqrt{1+\s}}.
\end{align}
Notice that for $A_{N-1}$ these central charges grow as $N^3$ in the large $N$ limit. Furthermore, they contain an infinite number of $1/N$ corrections. The order $N^3$ coefficient of these central charges was successfully matched to a supergravity computation \cite{Bah:2012dg}. In this paper we will extend this matching to the first subleading coefficient of order $N$. For this reason, it is useful to write the explicit form of the leading and subleading terms of the above expressions. For $\kappa =\pm 1$ we have:
\begin{align}
\label{eq:FTa4DorderNkappa1}	a &= |\mathfrak{g}-1|\Big(\frac{\kappa -9 \kappa  z^2+(3 z^2+1)^{3/2}}{48 z^2} N^3  -\frac{(z^2+1) (\kappa +\sqrt{3 z^2+1})}{16 z^2}N + \ldots\Big),\\
	c &= |\mathfrak{g}-1|\Big(\frac{\kappa -9 \kappa  z^2+(3 z^2+1)^{3/2}}{48 z^2} N^3 \nonumber\\
\label{eq:FTc4DorderNkappa1}	 & \qquad\qquad\qquad -\frac{z^2 (2 \sqrt{3 z^2+1}-\kappa)+3 (\kappa +\sqrt{3 z^2+1})}{48 z^2} N + \ldots\Big),
\end{align}
while the $\kappa = 0$ case leads to:
\begin{align}
\label{eq:FTa4DorderNkappa0}	a & = \frac{\sqrt{3} |z|}{16} N^3 - \frac{\sqrt{3}|z|}{16} N + \ldots,\\
\label{eq:FTc4DorderNkappa0}	c & = \frac{\sqrt{3} |z|}{16} N^3 - \frac{|z|}{8 \sqrt{3}} N + \ldots.
\end{align}
We will derive these results from a gravity computation in section \ref{sec:4DcentralchargesSUGRA}.
\subsection{Two Dimensions}\label{sec:2DCFT}

A two-dimensional analog of $a$-maximization was recently found by Benini and Bobev in \cite{Benini:2012cz,Benini:2013cda}. In two-dimensional theories with $(0,2)$ SUSY, there is a $U(1)_R$ associated with the right-movers. In general, the superconformal R-symmetry is related to the right-moving central charge $c_R$ by $c_R = 3 k^{RR}$, where $k^{RR}$ is the leading coefficient in the two-point function of the right-moving R-current. If there are additional Abelian global symmetries in the theory, then just as in four dimensions, the R-symmetry can appear to be ambiguous. The main result of \cite{Benini:2012cz,Benini:2013cda} is that the superconformal R-symmetry is determined by extremizing: 
\begin{equation}
\label{cext}
c_{R,t}(t_I) = 3 \left (k^{RR} + 2 \sum_I t_I k^{IR} + \sum_{IJ} t_I t_J k^{IJ} \right ).
\end{equation}
where the $k^{IJ}$ are the coefficients of the flavor current two-point functions (albeit, for left-movers, with an additional minus sign). Since $c_{R,t}$ is quadratic, it has a unique extremizing solution, which is a minimum for the directions corresponding to right-moving symmetries and maximum for the directions along the left-moving symmetries. Thus this procedure is simply called ``$c$-extremization".

Also, just as in four dimensions, we can consider the anomaly polynomial for a two-dimensional fermion charged under Abelian symmetries $F^I$. The anomaly four-form is given by:
\begin{equation}
I_4 = k^{IJ} c_1(F^I) c_1 (F^J) - \frac{k}{24}p_1(T)
\end{equation}
where $k^{IJ} = {\rm Tr} F^I F^J$ and $k = {\Tr}\, \gamma^3 = n_R - n_L$, the difference in the number of right-moving vs.~left-moving Weyl fermions. Just as in four dimensions, if we wrap M5-branes on a suitable four-dimensional space, we can integrate the 8-form over the compact space and read the appropriate charges off of $I_4$. These charges can then be used to do $c$-extremization.

The particular example we are going to study in the following is M5-branes wrapped on the product of two Riemann surfaces. In this case, the $c$-extremization procedure leads to the left and right central charges \cite{Benini:2012cz}:
\begin{align}
	c_R &= r_G\frac{\eta_1\eta_2}4\; \frac{\s^2 \cP + 3 \s \big( z_1^2 z_2^2 - 6\kappa_1\kappa_2 z_1z_2 + \kappa_1^2 \kappa_2^2 \big) - 9 \kappa_1 \kappa_2 z_1 z_2}{\s(\kappa_1 \kappa_2 - 3z_1 z_2) - 3 z_1 z_2}, \\
	c_L &=  r_G\frac{\eta_1\eta_2}4\; \frac{\s^2 \cP + 2 \s \big( 3 z_1^2 z_2^2 - 8\kappa_1\kappa_2 z_1z_2 + \kappa_1^2 \kappa_2^2 \big) + 3 z_1z_2( z_1z_2 - 2 \kappa_1 \kappa_2)}{\s(\kappa_1 \kappa_2 - 3z_1 z_2) - 3 z_1 z_2},
\end{align}
where we have defined:
\be \cP = 3z_1^2z_2^2+\kappa_1^2z_2^2+\kappa_2^2z_1^2-8\kappa_1\kappa_2z_1z_2+3\kappa_1^2\kappa_2^2,\ee
$\s=h_G(h_G+1)$ as before, and for each Riemann surface we define:
\be
\eta_i=\left\{\begin{matrix}1, && \mathfrak{g}_i=1, \\ 2|\mathfrak{g}_i-1|, && \mathfrak{g_i}\ne 1.\end{matrix}\right.
\ee
As before, it is convenient to expand the expressions above for $A_{N-1}$ to order $N$:
\begin{multline}
	\frac12(c_L + c_R) = \frac{\eta_1\eta_2}{4}\ls \frac{\cP}{\kappa_1\kappa_2-3z_1z_2}N^3\right.\\
	+ \frac{9z_1^3z_2^3+12\kappa_1^2z_1z_2^3+12\kappa_2^2z_1^3z_2+\kappa_1\kappa_2\lp 9z_1^2z_2^2-2z_1^2\kappa_2^2-2z_2^2\kappa_1^2+3\kappa_1\kappa_2z_1z_2-\kappa_1^2\kappa_2^2\rp}{2\lp\kappa_1\kappa_2-3z_1z_2\rp^2}N\\
\label{eq:FTcRpcL} \left.+\vphantom{\frac{\cP}{\kappa_1\kappa_2-3z_1z_2}} \ldots\rs
\end{multline}
\be
\label{eq:FTcLmcR}	c_R - c_L =  \frac{\eta_1\eta_2}4 (\kappa_1 \kappa_2 + z_1 z_2)N + \ldots
\ee
These results will be derived from a gravity computation in section \ref{sec:2DcentralchargesSUGRA}.

\section{$\cN=1$ SUGRA Review \& Solutions}
\label{sec:sugrarev}

In this section, we will discuss $\mathcal{N}=1$ supergravity in five dimensions with supersymmetric higher-derivative corrections, and construct supersymmetric $AdS_5$ and $AdS_3\times\Sigma_g$ solutions in this higher-derivative theory. The discussion of this supergravity theory takes place using the superconformal tensor calculus, but we will omit the details of the derivations of the actions and the gauge fixing that needs to happen to obtain Poincar\'e supergravity from the superconformal supergravity. Some more details of this gauge fixing (and references) are given in appendix \ref{app:confSUGRA}; however, since all that is needed to find the solutions we are interested in are the gauge-fixed actions and supersymmetry variations, these details can safely be skipped.

\subsection{Off-shell Multiplets \& Variations}
For computations involving higher-derivative corrections to supergravity theories, it is very useful to have an off-shell formulation of the supersymmetry multiplets, because the off-shell supersymmetry transformations are exact and do not receive any corrections at higher-derivative orders.  In contrast, if the supersymmetry algebra only closes on-shell, then as the equations of motion get modified by higher-derivative corrections, the supersymmetry transformations must also get corrected in order to maintain closure of the algebra.

In five dimensional $\cN=1$ supergravity, there is an off-shell superconformal formulation available for the Weyl multiplet containing the graviton as well as for vector multiplets.  Unfortunately, the theories we will obtain from the near-horizon limit of M5-branes wrapping a Riemann surface also have a hypermultiplet sector, and for these we only have an on-shell representation of the multiplet.\footnote{The off-shell formalism for hypermultiplets requires an infinite number of fields \cite{deWit:1984px}.}  However, we will find that having off-shell information even for only part of the theory will aid us in our analysis.

The off-shell standard Weyl multiplet contains the f\"unfbein $e_\m^a$, the gravitini $\psi_\m^i$,  and bosonic auxiliary fields $V_\m^{ij}$, $T_{ab}$, $D$, and $b_\m$, along with a fermion auxiliarino $\chi^i$.  Here, $\m$ and $a$, $b$ are curved and flat five-dimensional indices respectively, while $i$ and $j$ are doublet indices ($i,j=1,2$) of $SU(2)$.  $T_{ab}$ is an antisymmetric tensor, while $V_\m^{ij}$ is symmetric in its upper indices.  Our conventions for $SU(2)$ and for spinors are detailed in appendix \ref{app:SU2Conventions}. We will also have $n_V+1$ off-shell $\U(1)$ vector multiplets, each containing a gauge field $A_\m^I$ (with field strength $F^I_{\mu\nu}$), a real scalar $\rho^I$, a gaugino $\lambda^{I\,i}$, and an auxiliary field $Y^I_{ij}$ which is also a doublet of $SU(2)$. 

Finally, we will have $n_H$ on-shell physical hypermultiplets with scalars $q^{X}$ and fermions $\zeta^A$.  Here $X$ runs from $1$ to $4n_H$, while $A$ runs from $1$ to $2n_H$, so we have four real scalars and two $SU(2)$-Majorana fermions in each hypermultiplet. The superconformal tensor calculus will also require an extra, non-physical compensator hypermultiplet ~\cite{Bergshoeff:2004kh}. To get Poincar\'e supergravity in the superconformal formalism, we need to gauge fix the superconformal symmetries, which will entirely fix this compensator hypermultiplet (so that it disappears from the action) as well as set the Weyl multiplet field $b_\m=0$. We give more details on the superconformal fields and actions including compensators in appendix \ref{app:confSUGRA}, here we will always use the gauge-fixed variations and actions.

The $n_H+1$ hypermultiplets ({\it i.e.,} including the compensator) parametrize a hyperk\"ahler manifold, of which the physical hypermultiplets parametrize a quaternionic submanifold.\footnote{We refer to appendix \ref{app:hypers} for more details and references.} Important data of this manifold is given by the vielbein $f^{iA}_X(q)$ and associated quaternionic metric $h_{XY}(q)$ (both quantities are of the physical quaternionic manifold, so $X$ runs from $1$ to $4n_H$), and Killing vector parameters $k^X_I(q)$ which determine the charges of the physical hyperscalars under the gauge group. These parameters also determine associated moment maps $P_I^{ij}(q)$. More details on the hyperscalar quantities can be found in appendix \ref{app:hypers} (especially section \ref{sec:apphypersymmetries}).

In summary, the bosonic fields that survive the superconformal gauge fixing are given by: $e_\m^a$, $V_\m^{ij}$, $T_{ab}$, $D$; $A_\m^I$, $\rho^I$, $Y^I_{ij}$ ($I=1,\cdots,n_V+1$); $q^X$ ($X=1,\cdots,n_H$).

For bosonic solutions, all the gauge-fixed fermionic variations are:
\begin{align}
\nn \delta \psi^i_{\mu} 
&= \lp\partial_{\mu} + \frac14 \omega_{\mu}^{ab}\gamma_{ab}\rp\epsilon^i - V^{ij}_{\mu}\epsilon_j + i T^{ab}\lp\gamma_{ab}\gamma_{\mu}-\frac13\gamma_{\mu}\gamma_{ab}\rp\epsilon^i \non\\
\label{eq:vargravitino} & \quad + \frac{1}{3}\g_\m\g^a\Upsilon_a^{ij}\epsilon_j - \frac{i}{6} g \rho^I  \gamma_{\mu}P_I^{ij}\epsilon_j,\\
\delta\chi^i &= \frac{D}{4} \epsilon^i - \frac{1}{64} \gamma^{ab} \mathcal{F}^{ij}_{ab}\epsilon_j +\frac{i}{24}\g^{ab}\g^cT_{ab}\Upsilon_c^{ij}\e_j-\frac{1}{6}\g^{abcd}T_{ab}T_{cd}\e^i\non\\ \label{eq:varauxiliarino}& \quad +\frac{1}{24}g\rho^I\g^{ab}T_{ab}P_I^{ij}\e_j+\frac{i}{8}\g^{ab}\g^c\nabla_cT_{ab}\e^i-\frac{i}{8}\g^a\nabla^bT_{ab}\e^i \\
\nn \delta\lambda^{iI} 
&=  -\frac14 \gamma^{ab} F^I_{ab}\epsilon^i - \frac{i}{2} \g^a\partial_a\rho^I\epsilon^i-Y^{I\,ij}\epsilon_j + \frac{4}{3}\rho^I \gamma^{ab} T_{ab} \epsilon^i +\frac{i}{3}\rho^I\g^a \Upsilon_a^{ij}\epsilon_j + \frac{1}{6} g \rho^I\rho^J P_J^{ij}\epsilon_j   ,\\
\label{eq:varhyperino}\delta \zeta^A &= \frac{i}{2} (\partial_a q^X + g A^I_a k^X_I)f_X^{iA}\gamma^a\epsilon_i-\frac12 g \rho^I k_I^X f_{X}^{iA}\epsilon_i  ,
\end{align}
where we have defined:
\be
\nabla_\m T_{ab}=\p_\m T_{ab}-2\om_{\m[a}^{\hph{\m[a}c}T_{b]c},
\ee
and:
\begin{align}
\label{eq:mcFdef}
\mathcal{F}^{ij}_{\m\n}&=2\p_{[\m}V_{\n]}^{ij}-2V_{[\m}^{k(i}V_{\n]\,k}^{j)},\\
\Upsilon_a^{ij}&=V_a^{ij}-\hlf gA_a^IP_I^{ij}-\om_X^{ij}\p_aq^X,
\end{align}
where $\om_X^{ij}$ is the $SU(2)$ part of the spin connection on the hyperscalar manifold (see appendix \ref{app:confSUGRA}).

In the following, it will sometimes be useful to split objects with $SU(2)$ indices into a trace and a traceless part. We will denote the latter with a prime, so e.g.:
\be \label{eq:splitSU2trace} V^{ij}_a = \frac12 \delta^{ij} V_a + V'^{ij}.\ee 
Note that $SU(2)$ indices are raised and lowered with $\e_{ij}$ and $\e^{ij}$, not with $\d_{ij}$, in equations like (\ref{eq:mcFdef}).  Our conventions for $SU(2)$ indices are detailed in appendix \ref{app:SU2Conventions}.

\subsection{Two- and Four-Derivative Actions}
Here we will discuss the two- and four-derivative actions in the off-shell formalism. Once again, we only report the gauge-fixed Lagrangians; for more information about the gauge fixing, see appendix \ref{app:gaugefix}.

\subsubsection{Two Derivative Supergravity}
The two-derivative supergravity action is constructed in the superconformal formalism by first constructing the superconformal-invariant action and then gauge fixing to Poincar\'e supergravity. More details on the superconformal action is given in appendix \ref{app:confSUGRA}; here we will only state the gauge-fixed Poincar\'e supergravity Lagrangian. Note that the Weyl multiplet field $b_\m=0$ everywhere.

At two derivatives, the action is completely specified by a gauge coupling parameter $g$ and a cubic polynomial in the $\rho^I$:
\be
\mcC=C_{IJK}\rho^I\rho^J\rho^K,
\ee
where $C_{IJK}$ are totally symmetric constants.

Let us also define:
\be
\mcC_I=3C_{IJK}\rho^J\rho^K,\qquad\mcC_{IJ}=6C_{IJK}\rho^K,
\ee
and $(\mcC^{-1})^{IJ}$ is the matrix inverse to $\mcC_{IJ}$ (which we assume exists).

In terms of these, the two-derivative bosonic Lagrangian is:
\begin{align}
\nn e^{-1}\mathcal{L}_{R} &= \frac14 \mathcal{C}_{IJ} F^{I\,ab} F^J_{ab} +\frac12\mathcal{C}_{IJ} \partial_a\rho^I\partial^a\rho^J - \mathcal{C}_{IJ} Y_{ij}^I Y^{J\,ij} + 8(\mathcal{C}-1) D \\
\nn &+ (\frac{208}{3} \mathcal{C} - \frac{16}{3}) T^{ab}T_{ab} + (\frac14 \mathcal{C}+\frac{3}{4})R  - 8 \mathcal{C}_K F^K_{ab}T^{ab} +\frac{1}{4} e^{-1} \epsilon^{\mu\nu\rho\sigma\tau} C_{IJK} A^I_{\mu}F^J_{\nu\lambda}F^K_{\rho\sigma}\\
\nn & - h_{XY}  (\partial_a q^X + gA_a^I k_I^X)(\partial^a q^Y + gA^{Ia} k_I^Y)  + 2g Y^I_{ij} P_I^{ij} - g^2 \rho^I\rho^Jk^X_I k^Y_{J}h_{XY} \\
\label{eq:actionR} &+\hlf g^2\rho^I\rho^JP_I^{ij}P_{J\,ij} +2\Upsilon_a^{ij}\Upsilon^a_{ij},
\end{align}

The full equations of motion can be found in appendix \ref{app:twoderEOM}. Here, we mention that the equations of motion for the auxiliary fields $Y_{ij}^I$, $T_{ab}$, $V_a^{ij}$, and $D$ respectively give:
\begin{align}
 Y^I_{ij} &= g\lp\mcC^{-1}\rp^{IJ}P_{J\,ij},\\
T_{ab} &= \frac{1}{16}\mcC_IF^I_{ab},\\
\label{eq:REOMV}V_a^{ij} &= \hlf gA_a^IP_I^{ij}+\om_X^{ij}\p_aq^X \Longrightarrow\Upsilon_a^{ij}=0,\\
\label{eq:REOMD}\mcC&=1.
\end{align}

Note that $D$ is not determined by its own equation of motion, but can be determined by the other equations of motion. Rather, its equation of motion (\ref{eq:REOMD}) gives a constraint on the scalars that can be seen to give the correct normalization factor $1$ for the Einstein-Hilbert term in the action. Using these equations of motion, we can write the full two-derivative on-shell Lagrangian:
\begin{multline}
\label{eq:actionRonshell} 
e^{-1}\mathcal{L}_{R,on-shell} = R + \frac12 \mathcal{C}_{IJ} \partial_a\rho^I\partial^a\rho^J + \frac14 (\mathcal{C}_{IJ}-\mathcal{C}_I\mathcal{C}_J)F^{Iab}F^J_{ab} \\
+\frac{1}{4} e^{-1} \epsilon^{\mu\nu\rho\sigma\tau} C_{IJK} A^I_{\mu}F^J_{\nu\lambda}F^K_{\rho\sigma} - h_{XY}  (\partial_a q^X + gA_a^I k_I^X)(\partial^a q^Y + gA^{Ia} k_I^Y)  \\
+ g^2 (\mathcal{C}^{-1})^{IJ} P_{Iij} P_J^{ij} - g^2 \rho^I\rho^Jk^X_I k^Y_{J}h_{XY} +\frac12 g^2\rho^I\rho^JP^{ij}_I P_{ij,J},
\end{multline}
which has to be supplemented with the constraint $\mathcal{C}=1$, which should be solved before taking variations of (\ref{eq:actionRonshell}).

\subsubsection{Four Derivative Corrections}

We would now like to consider the higher-derivative corrections to this action.  There are three curvature-squared terms which could appear in principle, but we can remove one of them by a field redefinition.\footnote{In pure general relativity with no cosmological constant, we can remove two of the invariants by the two-parameter field redefinition
\be
g'_{\m\n}=g_{\m\n}+\ell_P^2\lp\alpha Rg_{\m\n}+\beta R_{\m\n}\rp.
\ee
However, in the given theory, a general redefinition of this form would also change Newton's constant by something proportional to the effective cosmological constant, essentially the potential for the scalars.  Since we choose to leave the Newton's constant fixed (though we will see in \eqref{eq:shiftGN} that the effective $G_N$ does get shifted), we are left with only a one-parameter field redefinition.}  We can choose to remove the Ricci squared term, leaving a Weyl squared term and a Ricci scalar squared term. The full supersymmetric action at four derivatives will be given by the supersymmetric completions of these two four-derivative curvature terms.\footnote{In principle, one could wonder whether there are any other four-derivative Lagrangians that contain, say, higher derivatives in the hyperscalars only. However, since for all our solutions we are only interested in hyperscalars that are constant and moreover covariantly constant, any such possible Lagrangians could not contribute to our results.} These two actions will introduce two new sets of parameters $c_I, b_I$, which should be seen as small parameters compared to the curvature of solutions in order for the derivative expansion to be meaningful.

The supersymmetric completion of the Weyl squared term was first calculated in \cite{Hanaki:2006pj} and contains a mixed gauge-gravitational Chern-Simons term $\sim A\wedge R\wedge R$. The bosonic part of the action is given by \cite{Hanaki:2006pj,Ozkan:2013nwa}:
\begin{align}
\nn \mathcal{L}_{C^2}&=\sqrt{-g} c_I\left\{\frac{1}{8}\rho^IC^{\m\n\rho\s}C_{\m\n\rho\s}+\frac{64}{3}\rho^ID^2+\frac{1024}{9}\rho^IT^2D-\frac{32}{3}DT_{\m\n}F^{I\,\m\n}\right.\nn\\
\nn &\left.-\frac{16}{3}\rho^IC^{\m\n\rho\s}T_{\m\n}T_{\rho\s}+2C^{\m\n\rho\s}T_{\m\n}F^I_{\rho\s}+\frac{1}{16}\e^{\m\n\rho\s\lambda}A^I_\m C_{\n\rho}^{\hph{\n\rho}\tau\d}C_{\s\lambda\tau\d}-\frac{1}{12}\e^{\m\n\rho\s\lambda}A^I_\m \mathcal{F}_{\n\rho}^{ij}\mathcal{F}_{\s\lambda\,ij}\right.\nn\\
\nn &\left.+\frac{16}{3}Y^I_{ij}\mathcal{F}_{\m\n}^{ij}T^{\m\n}-\frac{1}{3}\rho^I\mathcal{F}_{\m\n}^{ij}\mathcal{F}^{\m\n}_{ij}+\frac{64}{3}\rho^I\nabla_\n T_{\m\rho}\nabla^\m T^{\n\rho}-\frac{128}{3}\rho^IT_{\m\n}\nabla^\n\nabla_\rho T^{\m\rho}\right.\nn\\
\nn &\left.-\frac{256}{9}\rho^IR^{\n\rho}T_{\m\n}T^\m_{\hph{\m}\rho}+\frac{32}{9}\rho^IRT^{\m\n}T_{\m\n}-\frac{64}{3}\rho^I\nabla_\m T_{\n\rho}\nabla^\m T^{\n\rho}+1024\rho^IT^{\m\n}T_\m^{\hph{\m}\rho}T_\n^{\hph{\n}\s}T_{\rho\s}\right.\nn\\
\nn &\left.-\frac{2816}{27}\rho^IT^{\m\n}T_{\m\n}T^{\rho\s}T_{\rho\s}-\frac{64}{9}T^{\m\n}T_{\m\n}T^{\rho\s}F^I_{\rho\s}-\frac{256}{3}T^{\m\n}T_\m^{\hph{\m}\rho}T_\n^{\hph{\n}\s}F^I_{\rho\s}\right.\nn\\
\label{eq:actionC2}&\left.-\frac{32}{3}\e^{\m\n\rho\s\lambda}T_\m^{\hph{\m}\tau}\nabla_\tau T_{\n\rho}F^I_{\s\lambda}-16\e^{\m\n\rho\s\lambda}T_\m^{\hph{\m}\tau}\nabla_\n T_{\rho\tau}F^I_{\s\lambda}-\frac{128}{3}\rho^I\e^{\m\n\rho\s\lambda}T_{\m\n}T_{\rho\s}\nabla^\tau T_{\lambda\tau}\right\},
\end{align}
where the Weyl tensor in five dimensions is:
\be
C_{\m\n\rho\s}=R_{\m\n\rho\s}-\frac{1}{3}\lp g_{\m\rho}R_{\n\s}-g_{\n\rho}R_{\m\s}-g_{\m\s}R_{\n\rho}+g_{\n\s}R_{\m\rho}\rp+\frac{1}{12}\lp g_{\m\rho}g_{\n\s}-g_{\m\s}g_{\n\rho}\rp R.
\ee

The supersymmetric completion of the Ricci scalar squared term was computed in \cite{Ozkan:2013nwa} using the superconformal formalism, but using a linear multiplet compensator. This linear multiplet, together with the Weyl multiplet, can then be embedded in a composite vector multiplet that is needed for the construction of the Ricci scalar squared action. Since we use a hypermultiplet compensator instead of a linear one, we need to map our hyperscalar compensator multiplet into a linear multiplet.\footnote{Note that the Weyl squared Lagrangian is completely independent of which compensator is used, so such a translation between compensator multiplets is never necessary there.} We can do this using the formulae in \cite{Fujita:2001kv} for embedding hyperscalar multiplets into linear multiplets. Finally, the bosonic parts of the composite vector multiplet needed for the action is given by (after gauge fixing):
\begin{align}
\ul{\rho} &= \frac{1}{\sqrt{2}} g \rho^IP_I,\\
\nn \ul{Y}^{ij} &= \frac{1}{\sqrt{2}}\delta^{ij}\left( -\frac38 R - \frac18 (g\rho^IP_I)^2  - \frac18 \Upsilon^2 +\frac83 T^2 + 4 D - V'^{kl}_aV'^a_{kl}\right)\\
& \qquad + \frac{1}{\sqrt{2}}\Upsilon^aV'^{ij}_a - \sqrt{2}\nabla^a V'^{ij}_a,\\
\ul{F}^{ab} &= 2\sqrt{2}\partial^{[a}\left( V^{b]} + \frac12\Upsilon^{b]}\right),
\end{align}
where we have used the trace and traceless parts of $V^{ij}_a,\Upsilon^{ij}_a,P_I^{ij}$ as in (\ref{eq:splitSU2trace}). In terms of this composite vector multiplet and the other fields, the bosonic part of the supersymmetric Ricci scalar squared action is given by \cite{Ozkan:2013nwa}:
\begin{align}
\nn\mathcal{L}_{R^2} &= \sqrt{-g}b_I\left\{ \rho^I \ul{Y}_{ij}\ul{Y}^{ij} + 2\ul{\rho}\ul{Y}^{ij} Y^I_{ij} - \frac18 \rho^I\ul{\rho}^2 R -\frac14
\rho^I\ul{F}_{\mu\nu}\ul{F}^{\mu\nu}-\frac12\ul{\rho}\ul{F}^{\mu\nu}F_{\mu\nu}^I\right.\\
\nn & \left. + \frac12\rho^I\partial_\m\ul{\rho}\partial^\m \ul{\rho} + \rho^I\ul{\rho}\nabla^2\ul{\rho}-4\rho^I\ul{\rho}^2(D+\frac{26}{3}T^2) + 4\ul{\rho}^2F_{\mu\nu}^IT^{\mu\nu}\right.\\
\label{eq:actionR2} &\left. + 8 \rho^I \ul{\rho}\ul{F}_{\mu\nu}T^{\mu\nu} - \frac18 \epsilon_{\mu\nu\rho\sigma\lambda}A^{\mu I} \ul{F}^{\nu\rho} \ul{F}^{\sigma\lambda}\right\}.
\end{align}

The total Lagrangian is then given by $\mathcal{L}_R+\mathcal{L}_{C^2}+\mathcal{L}_{R^2}$; the equations of motion for the auxiliary fields $D$ and (the trace part of) $Y_{ij}^I$ following from this action give:
\begin{multline}
 \mathcal{C} = 1  -c_I\left(\frac{16}{3} \rho^I D + \frac{1}{18} \rho^I (\mathcal{C}_JF^J)^2 - \frac{1}{12} \mathcal{C}_J F^J\cdot F^I\right)\\
\label{eq:DEOM1}  - b_I\left(4\rho^I D + \frac12 g \rho^JP_J Y^I - \frac38 R\rho^I - \frac38 \rho^I g^2(\rho^JP_J)^2+\frac{1}{96}\rho^I (\mathcal{C}_JF^J)^2\right),
\end{multline}
\begin{multline}
 Y^I = (\mathcal{C}^{-1})^{IJ} gP_J + (\mathcal{C}^{-1})^{IJ}c_J \frac{1}{12} g P_K F^K\cdot F^L\mathcal{C}_L\\
\label{eq:YEOM1} + (\mathcal{C}^{-1})^{IJ}b_J (g \rho^KP_K)\left(-\frac38 R - \frac18 g^2 (\rho^LP_L)^2 + \frac{1}{96}(\mathcal{C}_LF^L)^2 + 4D\right),
\end{multline}
where we have used the leading order (two-derivative) equation of motion for $T$ and $V$ to simplify the higher-order piece. The auxiliary equation of motion for $V$ and $T$ are quite a bit more complicated, so we omit them here.

\subsection{Supersymmetric Solutions}
\label{sec:susysols}
Now, we can discuss finding $AdS_5$ and $AdS_3\times\Sigma_g$ solutions in our $\mathcal{N}=1$ supergravity theory with higher-derivative corrections. For our solutions, we will always take constant hyperscalars (which implies constant $k^X_I,P_I^{ij}$), and moreover we will demand that:
\begin{align}
\label{eq:kIrhoI0} k^X_I\rho^I &= 0,\\
\label{eq:kIAI0} k^X_I A^I_{\mu} &= 0. 
\end{align}
These conditions will clearly set the hyperino variation (\ref{eq:varhyperino}) identically to zero. See appendix \ref{app:hyperinovar} for more discussion on the hyperino variation and the conditions (\ref{eq:kIrhoI0}) and (\ref{eq:kIAI0}).

We will also take a diagonal $SU(2)$ ansatz, which is an ansatz often used to find solutions to this theory. This ansatz consists of taking all fields in the Weyl and vector multiplets that have symmetric $SU(2)$ indices to only consist of the trace part, i.e. $V'^{ij}=P'^{ij}_I=Y'^{ijI}=0$ in the notation of (\ref{eq:splitSU2trace}). See also appendix \ref{app:su2vars} for more information. Note that this is actually a restriction on the allowed possible $P^{ij}_I$ and thus a restriction on the hypermultiplets.

When we take this ansatz, the supersymmetry variations can be seen to simplify to (again, see appendix \ref{app:su2vars} for more discussion on the simplification of the variations):
\begin{align}
\label{eq:simplvarpsi}\delta\psi_{\mu} &= \left( \partial_{\mu} + \frac14 \omega_{\mu}^{ab} \gamma_{ab} - \frac{i}{2} V_{\mu} + \frac{i}{12}\gamma_{\mu} \slashed{\Upsilon} + i T^{ab}(\gamma_{ab}\gamma_{\mu}-\frac13\gamma_{\mu}\gamma_{ab}) + \frac{1}{12} g\rho^I P_I \gamma_{\mu}\right)\epsilon,\\
\nn \delta\chi &= \left( \frac14 D - \frac{i}{64} \gamma^{ab}\partial_aV_b + \frac{i}{8} \gamma^{ab} \slashed{\nabla} T_{ab} - \frac{i}{8} \gamma^a \nabla^b T_{ab} + \frac16 T^2 - \frac14 \gamma^{abcd}T_{ab}T_{cd}\right.\\
\label{eq:simplvarchi}& \left. + \frac{1}{12}(\gamma\cdot T)^2 + \frac{i}{48} \gamma\cdot T g \rho^IP_I  -\frac{1}{48} \gamma\cdot T \slashed{\Upsilon}\right)\epsilon,\\
\label{eq:simplvarlambda}\delta\lambda^I &= \left(-\frac14 \gamma\cdot F^I - \frac{i}{2} \slashed{\partial}\rho^I - \frac{i}{2} Y^I + \frac{4}{3} \rho^I \gamma\cdot T +\frac{i}{12} g \rho^I \rho^J P_J  -\frac{1}{12}  \rho^I \slashed{\Upsilon}\right) \epsilon.
\end{align}
where we have defined $\epsilon = \epsilon^1 + i \epsilon^2$ and similar for the other spinors involved. We do not need to consider the hyperino variations anymore as we have set them to zero by demanding (\ref{eq:kIrhoI0}) and (\ref{eq:kIAI0}).

\subsubsection{Supersymmetric $AdS_5$ Solutions}\label{sec:AdS5sol}
To find (maximally) supersymmetric $AdS_5$ solutions, we set all spin $>0$ fields to zero (i.e. $V_a=T_{ab}=F_{ab}=0$, which also automatically implies $\Upsilon_a=0$ since we have constant hyperscalars) and set all scalars to be constants. We take the metric to be:
\be ds^2 = \frac{L^2}{r^2}\left(-dt^2 + dr^2 + dx^2+dy^2+dz^2\right),\ee
which has Ricci scalar $R=-20L^{-2}$.

The variations (\ref{eq:simplvarchi}) and (\ref{eq:simplvarlambda}) respectively give:
\begin{align}
 D &= 0,\\
\label{eq:AdS5solY} Y^I &= \frac16 \rho^I (g\rho^J P_J).
\end{align}
Additionally, the variation (\ref{eq:simplvarpsi}) must read:
\be \delta\psi_{\mu} = D_{\mu}\epsilon + \frac{1}{2 L}\epsilon,\ee
in order for there to be eight linearly independent solutions $\epsilon$ to $\delta\psi_{\mu}=0$, and thus for the solution to preserve maximum supersymmetry. This fixes the radius of AdS to be:
\be L^{-1} = \frac16 (g\rho^IP_I).\ee

We can use the equations of motion (\ref{eq:DEOM1}) and (\ref{eq:YEOM1}) with our ansatz to get:
\begin{align}
\label{eq:AdS5Ceq} \mathcal{C} &= 1 + 3b_I \rho^I L^{-2},\\
\label{eq:AdS5CIeq} \mathcal{C}_I =\pi_I& \equiv (g P_I) \frac{L}{2} + 9 b_I L^{-2}.
\end{align}
This can be solved exactly when the vector-scalar manifold $\cM_v$ is a homogeneous space with metric $g_{IJ} = \mathcal{C}_I\mathcal{C}_J-\mathcal{C}_{IJ}$, since in this case the following tensor has constant entries:
\begin{align}
\widehat{\mathcal{C}}^{IJK}&=\frac{1}{\mathcal{C}^2} g^{IL} g^{JM} g^{KN} C_{LMN},
\end{align}
and obeys the following identity:
\begin{align}
\widehat{\mathcal{C}}^{IJK} C_{J(LM}C_{NP)K}&=\frac{1}{27}\delta^{I}_{(L}C_{MNP)}\,.
\end{align}
Then the solution to (\ref{eq:AdS5CIeq}) is:
\begin{align}
\rho^I =3 \frac{\widehat{\mathcal{C}}^{IJK}\pi_J \pi _K}{\sqrt{\widehat{\mathcal{C}}^{IJK}\pi_I \pi_J \pi_K} }
\end{align}
The radius $L$ is given by:
\begin{align}
\label{eq:AdS5Lsol}L^{-1} &= \frac12 P\left(1 + \frac14 b_P\right),
\end{align}
where we have defined:
\begin{align}
\label{eq:defP} P &=  (g^3 \widehat{\mathcal{C}}^{IJK} P_I P_J P_K)^{1/3},\\
\label{eq:defbP}b_P &= 3 g^2 \widehat{\mathcal{C}}^{IJK}b_IP_JP_K. 
\end{align}

As an example we now present the STU-model more explicitly: we have three vector multiplets and the symmetric tensor $C_{IJK}$ has $C_{123}=1/6$ and permutations thereof, with other components vanishing. In this case we have non-vanishing components $\widehat{\mathcal{C}}^{123} = 1/6$ and permutations, and we can write the solution for the scalar fields explicitly as:
\begin{align}
\label{eq:AdS5rhosol}\rho^1 &= \frac{P}{gP_1}\left(1 + g^2(b_2P_1P_3+b_3P_1P_2 - \frac54 b_1P_2P_3) \right), \qquad \text{(similarly for $\rho^2,\rho^3$)}
\end{align}

Finally, we note that (\ref{eq:kIAI0}) is automatically satisfied but (\ref{eq:kIrhoI0}) is not. We should see (\ref{eq:kIrhoI0}) as determining the constant values for the hyperscalars as follows: $P_I(q)$ is a function of the hyperscalars, so $\rho^I(q)$ are as well (due to (\ref{eq:AdS5rhosol})). Then (\ref{eq:kIrhoI0}) gives us the equation $k_I^X \rho^I(q) = 0$, which should be thought of as equations determining the possible (constant) values for the hyperscalars $q^X$. We will see an explicit example of this below in section \ref{sec:4DcentralchargesSUGRA} once we specify the specific hyperscalar manifold and $k_I^X, P_I$ for our M5-brane system.

\subsubsection{Supersymmetric $AdS_3\times\Sigma_g$ solutions}\label{sec:AdS3sol}
We can follow the same procedure to find (quarter-)supersymmetric $AdS_3\times \Sigma_g$ solutions to our theory: first, we consider the off-shell supersymmetric variations on our ansatz; then, we use a few of the simpler auxiliary field equations of motion to fully determine the solution. This procedure is very similar to that used in \cite{Castro:2007sd,Castro:2007hc} to find $AdS_3\times S^2$ solutions in higher-derivative ungauged supergravity.

Our metric ansatz is taken to be:
\be ds^2 = \frac{e^{2f_0}}{r^2}(-dt^2 + dz^2 + dr^2) + e^{2g_0+2h(x,y)} (dx^2 + dy^2),\ee
with $f_0,g_0$ constants and $h(x,y)$ the metric function of the Riemann surface of genus $\mathfrak{g}$ spanned by $(x,y)$:
\be
h(x,y) = \begin{cases}
    -\log \frac{1+x^2+y^2}{2}, & \text{for $\mathfrak{g}=0$},\\
\frac12 \log 2\pi, & \text{for $\mathfrak{g}=1$},\\
-\log y, & \text{for $\mathfrak{g}>1$}.
  \end{cases}  
\ee
Note that these satisfy:
\be e^{-2h}(\partial_x^2+\partial_y^2)h = -\kappa,\ee
where $\kappa$ is $1$, $0$, or $-1$ for $\mathfrak{g}=0$, $\mathfrak{g}=1$, or $\mathfrak{g}>1$ respectively.

We will use the obvious choice for vielbeins:
\begin{align}
 e^{\hat{\mu}} &= \frac{1}{r} e^{f_0} dx^{\mu}, & (\mu=t,z,r)\\
 e^{\hat{\mu}} &= e^{g_0+h} dx^{\mu}, & (\mu=x,y).
\end{align}

All scalars are taken to be constants.  We take the components of the gauge fields $F^I$ and the auxiliary field $T$ along $\hx\hy$ to be their only non-zero components. We note that for such an ansatz, clearly $F\wedge F = 0$. We can also see that the equation of motion for $V$ at higher-derivative order simplifies to:
\be \label{eq:AdS3VEOM} \Upsilon_{\mu} = 0,\ee
in other words, the higher-derivative terms do not contribute to this equation of motion. We will use this to immediately set $\Upsilon_{\mu}=0$ in the supersymmetry variations.

We also impose the projection conditions (effectively killing all but one-quarter of the supersymmetries) on the Killing spinor:
\begin{align} 
\gamma_{\hr}\epsilon &= \epsilon,\\
\gamma_{\hx\hy}\epsilon&= i \epsilon.
\end{align}
Further, we take the Killing spinor $\epsilon$ to only depend on $r$.

The gravitino variation (\ref{eq:simplvarpsi}) along $\hatt,\hz$ sets $V_{\hz}=V_{\hatt}=0$ and in addition gives us:
\be \label{eq:gen1}\frac32 \omega_{\hz}^{\ \hz \hr} - 4T_{\hx\hy} + \frac{1}{4}g\rho^IP_I =0 .\ee
Similarly, the gravitino variation along $i=\hx,\hy$ gives $V_i= \omega_i^{\hx\hy}$ and:
\be  \label{eq:gen2}  8 T_{\hx\hy} + \frac{1}{4}g\rho^IP_I =0.\ee
Finally, the $\hr$ component of the variation gives $V_{\hr}=0$ and:
\be \label{eq:gen4} \partial_{\hr}\epsilon = \left(\frac43 T_{\hx\hy} - \frac{1}{12}g\rho^IP_I\right)\epsilon.\ee
The gaugino variation (\ref{eq:simplvarlambda}) gives us:
\be \label{eq:gen5} Y^I = - F_{\hx\hy}^I + \frac{16}{3}\rho^I T_{\hx\hy} + \frac{1}{6}g\rho^I\rho^JP_J.\ee
Finally, the auxiliarino variation (\ref{eq:simplvarchi}) gives us:
\be \label{eq:gen6} D = -\frac18 \partial_{[\hx}V_{\hy]} + \frac16 g\rho^IP_I  T_{\hx\hy} + \partial_{\hr} T_{\hx\hy}.\ee
These SUSY variations can be seen to completely determine the auxiliary fields as well as provide a relationship between the $AdS_3$ radius parametrized by $f_0$ and the scalars $\rho^I$:
\begin{align}
g \rho^I P_I &= 4 e^{-f_0},\\
T_{\hx\hy} &= -\frac18 e^{-f_0},\\
\label{eq:AdS3solY} Y^I &= -F_{\hx\hy}^I,\\
D &= -\frac{1}{16} e^{-2g_0} \kappa - \frac{1}{12} e^{-2f_0},\\
\mathcal{F}_{\hx \hy} &= e^{-2g_0}\kappa.
\end{align}

Now, we turn to the equations of motion for the auxiliary fields $V,Y,D$. First of all, the equation of motion for $V$ was discussed above in (\ref{eq:AdS3VEOM}) and gives us:
\be \frac12 e^{2g_0} g F^I_{\hx\hy} P_I = \kappa.\ee
We see that $F^I_{\hx\hy}$ should be constant, which we parametrize by:
\be F^I_{\hx\hy} = - a^I e^{-2g_0},\ee
so that the equation of motion for $V$ reduces to:
\be \label{eq:fin1} \frac12 g a^I P_I = -\kappa.\ee

Next, we take the equations of motion for $D$ (\ref{eq:DEOM1}) and $Y^I$ (\ref{eq:YEOM1}). These simplify to:
\begin{align}
\nn \mathcal{C} = 1 &- \frac{c_I}{6}e^{-2g_0}gP_J\left(\rho^Ia^J-\frac12 a^I\rho^J\right)\\
&-b_I\left(2a^I e^{-2g_0}e^{-f_0} - \rho^I e^{-2g_0}\kappa - 4 e^{-2f_0}\rho^I\right).\\
\nn a^I e^{-2g_0} = (\mathcal{C}^{-1})^{IJ}gP_J &-\frac23(\mathcal{C}^{-1})^{IJ}c_J \kappa e^{-f_0}e^{-2g_0}\\
&- 4(\mathcal{C}^{-1})^{IJ}b_J \kappa e^{-f_0}e^{-2g_0}.
\end{align}
These equations, together with the constraint $g \rho^I P_I = 4e^{-f_0}$ found above, fully determine the $AdS_3$ solution $\rho^I, f_0,g_0$. We can explicitly solve these equations in general. We first define:
\be \mathcal{C}_{IJ} a^J = \tilde{\mathcal{C}}_{IJ}\rho^J,\ee
i.e. $\tilde{\mathcal{C}}_{IJ}=6C_{IJK}a^K$. We assume this matrix is invertible, with $\tilde{\mathcal{C}}^{IJ}\equiv (\tilde{\mathcal{C}}^{-1})^{IJ}$. We define the shorthands:
\begin{align}
K_{I,1} &= -\frac23 c_I\kappa - 4b_I \kappa,\\
L &= g^2\tilde{\mathcal{C}}^{IJ}P_IP_J,\\
M_1 &= g\tilde{\mathcal{C}}^{IJ} P_I K_{J,1},\\
N &= g^3C_{IJK} \tilde{\mathcal{C}}^{IJ} \tilde{\mathcal{C}}^{JM} \tilde{\mathcal{C}}^{KN} P_L P_M P_N,\\
\tilde{N}_1 &= 3g^2 C_{IJK} \tilde{\mathcal{C}}^{IL} \tilde{\mathcal{C}}^{JM} \tilde{\mathcal{C}}^{KN} P_L P_M K_{N,1},
\end{align}
where the sub/superscript $1$ denotes that a quantity is at subleading order (so is proportional to $c_I,b_I$). Then the general solutions can be found for $f_0,g_0,\rho^I$:
\begin{align}
\label{eq:AdS3rhosol} \rho^I &= g\tilde{\mathcal{C}}^{IJ} P_J e^{2g_0} + \tilde{\mathcal{C}}^{IJ} K_{J,1} e^{-f_0},\\
e^{-f_0} &= \frac14 L e^{2g_0}\left(1+\frac{1}{4} M_1\right),\\
e^{6g_0} &= \frac{1}{N}\left( 1 + \frac12 L a^I \left(\frac16c_I - b_I\right) + g\kappa\tilde{\mathcal{C}}^{IJ}P_J\left(\frac13c_I+b_I\right) - \frac{\tilde{N}_1 L}{4N} + \frac{1}{4} \frac{L^2g \tilde{\mathcal{C}}^{IJ}P_Ib_J}{N}\right).
\end{align}

For the STU model, with $C_{123}=1/6$ as the only components, at leading order ($c_I=b_I=0$) these solutions simplify to \cite{Benini:2013cda}:
\begin{align}
(e^{3f_0})_{(0)} &= -\frac{8 a^1 a^2 a^3 \Pi}{\Theta^3},\\
(e^{6g_0})_{(0)} &= \frac{(a^1 a^2 a^3)^2}{\Pi},\\
(\rho^1)^3_{(0)} &= \frac{(a^1)^2}{a^2a^3}\frac{A_1^2}{A_2A_3}, \qquad \text{(similar for $\rho^2,\rho^3$)}.
\end{align}
The solutions are given in terms of the following combinations of $a^I, g P_I$:
\begin{align}
 \Pi &= A_1A_2A_3,\\
 \Theta & = \frac12\left(\tilde{A}_1+\tilde{A}_2+\tilde{A}_3\right),\\
 A_1 &= \frac{g}{2}(-a^1P_1 + a^2P_2 + a^3P_3) = -\kappa-ga^1P_1 \qquad \text{(similar for $A_2,A_3$)},\\
  \tilde{A}_1 &= \left(\frac{g}{2}\right)^2 \left((a^2P_2)^2+(a^3P_3)^2-a^1P_1(a^2P_2+a^3P_3)-2a^2P_2a^3P_3\right) \text{(similar for $\tilde{A}_2,\tilde{A}_3$)},
\end{align}
We stress that we will not actually need the higher-derivative corrections to the $AdS_3$ background (as opposed to the situation in $AdS_5$ above) to calculate the central charges, as we will see when we specify the M5 brane system.

We note that both (\ref{eq:kIAI0}) and (\ref{eq:kIrhoI0}) still need to be satisfied. As in the $AdS_5$ case, we can view (\ref{eq:kIrhoI0}) as fixing the (constant) hyperscalars. The equation (\ref{eq:kIAI0}) represents a real restriction on the possible gauge field strengths given by $a^I$ that preserve supersymmetry. These constants $a^I$ thus must satisfy:
\begin{align} 
\frac12 g a^I P_I &= -\kappa,\\
 k_I^X a^I &= 0.
\end{align}
These are two non-trivial relations the $a^I$ must satisfy. In the M5-brane solution below in section \ref{sec:2DcentralchargesSUGRA}, we will have three parameters $(a^1,a^2,a^3)$; we can then view $k_I^X a^I = 0$ as fixing $a^3$ in terms of $a^1,a^2$, and view $g a^I P_I/2 = -\kappa$ as allowing us to parametrize $a^1,a^2$ in terms of $\kappa$ and a free parameter $z$.

\section{M5-branes wrapped on Riemann surfaces}
\label{sec:hyperscalars}
In this section we determine the specific form of the actions that describe M5-branes wrapped on one or two Riemann surfaces. Our strategy will be as follows:
\begin{itemize}
	\item We determine the geometry of the scalars in the vector and hyper multiplets. The former was determined in \cite{Szepietowski:2012tb}, and we will argue that the latter is described by $SU(1,2)/U(2)$.
	\item Using the results of \cite{Harvey:1998bx} we derive the form of (some of) the Chern-Simons terms appearing in the 5d effective action at subleading order.
	\item We fix the form of the four-derivative terms by completing the Chern-Simons terms to supersymmetric invariant structures.
\end{itemize}
This allows us to derive the main results of this paper, the subleading corrections to $a$ and $c$ for M5-branes wrapped on one Riemann surface, and $c_L$ and $c_R$ for M5-branes wrapped on two Riemann surfaces. When compared to the field theory results derived using $a$-maximization \cite{Bah:2011vv} and $c$-extremization \cite{Benini:2012cz} respectively, we find complete agreement.

\subsection{The scalar geometry at two derivatives}
We begin by briefly reviewing the reduction of 7d $U(1)^2$ gauged supergravity on a Riemann surface $\Sigma_2$ of genus $\mathfrak{g}_2$, as presented in \cite{Szepietowski:2012tb}. This will immediately give us the scalar geometry for the vector multiplet. As explained in \cite{Szepietowski:2012tb}, only two of the four hyperscalars are kept in this truncation, so the geometry of the scalars in the hypermultiplet is not immediately visible from the resulting 5d action. We will show that we can introduce two additional hyperscalars that, together with the two retained in the compactification, parametrize the quaternionic manifold $SU(1,2)/U(2)$. Even though the final results will not be affected by the precise quaternionic geometry in the hypermultiplet sector, this allows us to make the discussion more concrete. Some details and references on the 7d $U(1)^2$ gauged supergravity can be found in appendix \ref{app:7Dgauged}. For the purposes of defining some useful symbols, however, we note here that the truncation we consider contains a gauge coupling $m$, and its field content has two scalars $\lambda_1,\lambda_2$, two $U(1)$ vector fields $F^{(1)},F^{(2)}$, and a three-form potential $S$.

We take the metric, field strength, and 3-form to be:
\begin{align}
ds_7^2 &= e^{-\frac{4B}{3}} ds_5^2 + e^{2B} ds_{\Sigma_2}^2,\\
\label{eq:7DreddefF}F^{(i)} &= \frac12 F^I + \frac14 p_i vol_{\Sigma_2},\\
S &= c_3 + c_1\wedge vol_{\Sigma_2},
\end{align}
where the 7d index $i$ runs over $i\in\{1,2\}$. The equation of motion for the 3-form $S$ in 7d leads to the constraint:
\begin{align} c_3 = -\frac{1}{m} e^{-8B/3+4\lambda_1+\lambda_2} *_5\left[ dc_1 + \frac{1}{\sqrt{3}m} (p_1 F^1 + p_2 F^2)\right],\end{align}
which allows us to eliminate the 3-form in 7d. If we want to have diagonal kinetic terms for the three vectors thus obtained, we can use the basis $(A^3,A^1,A^2)$, where:
\begin{align} A^3 = -\sqrt{3} \left(2c_1 - \frac{2}{\sqrt{3} m} (p_2 A^1 + p_1 A^2)\right).\end{align}
We also note that we can parametrize $p_A$, where $A\in\{1,2\}$, as:
\be \label{eq:defpi} p_1 = -\frac{\kappa_2-z_2}{m},\qquad p_2 = -\frac{\kappa_2+z_2}{m},\ee
so that $m(p_1+p_2)/2=-\kappa_2$, as required by 7d SUSY. 

If we define:
\begin{align} X^3 = e^{\frac{4B}{3}-2\lambda_1-2\lambda_2}, \qquad X^1 = e^{-\frac{2B}{3}+2\lambda_1}, \qquad X^2 = e^{-\frac{2B}{3}+2\lambda_2},\qquad H = e^{B+\lambda_1+\lambda_2},\end{align}
the resulting 5d Lagrangian reads:
\begin{align} 
\mathcal{L}_5 &= R *1 - \frac12\sum_I \frac{1}{(X^I)^2} F^I\wedge *F^I - \frac12 \sum_I \frac{1}{(X^I)^2} dX^I\wedge *dX^I - 2\frac{1}{H^2} dH\wedge *dH\nn\\
& + A^1\wedge F^2\wedge F^3 - V *1\nn\\
& -  \left(\frac{m^2}{2H^4} A^3\wedge *A^3 + \frac{p_2^2}{8H^4} A^1\wedge * A^1 + \frac{p_1^2}{8H^4} A^2\wedge *A^2\right.\nn\\
\label{eqlagrPS} &\left. + \frac{p_1p_2}{4H^4} A^1\wedge *A^2 - \frac{m}{2H^4} A^3\wedge *(p_2 A^1 + p_1 A^2)\right),
\end{align}
where the potential $V$ is given by:
\begin{align}
V &= -\frac{2m^2}{H^2}\left(\frac{1}{X^1} + \frac{1}{X^2}\right) - \left(4m^2-m\frac{p_1+p_2}{H^2}\right)\left(\frac{1}{X^3}\right) \nn\\
& + \frac{p_2^2}{8H^4} (X^1)^2 + \frac{p_1^2}{8H^4} (X^2)^2 + \frac{m^2}{2H^4} (X^3)^2.
\label{thelagrPS}
\end{align}

It is immediately obvious that in this basis, the geometry of the scalars in the vector multiplets is given by a symmetric rank three tensor $C_{IJK}$ whose only non-zero components are $C_{123}=1/6$ and permutations thereof. However, as pointed out in \cite{Szepietowski:2012tb}, this Lagrangian retains only one hyperscalar while a full hypermultiplet contains four. We will now argue that the hyperscalars can be chosen to parametrize the quaternionic manifold $SU(2,1)/U(2)$, and we will identify the isometries that are gauged.

\subsubsection*{$SU(2,1)/U(2)$ geometry}
We parametrize $SU(2,1)/U(2)$ by the coordinates $q^X = (\rho,\psi,x,y)$. A brief review of quaternionic geometry is presented in appendix \ref{app:hyperquaternionic}. The metric is given by:
\begin{equation}
	\label{eq:su21u2metric}
	ds^2 = \frac{d\rho^2}{2\rho^2} + 8 \frac{1}{\rho^2} \left(d\psi + \frac12 \left(y dx - x dy\right) \right)^2 + 2 \frac{1}{\rho}\left(dx^2 + dy^2\right).
\end{equation}
It is apparent from \eqref{eq:su21u2metric} that the metric has translational symmetry in the coordinate $\psi$ and rotational symmetry in the $(x,y)$ plane.\footnote{For a more complete treatment of the isometries, see \cite{Behrndt:2001km}.} The corresponding Killing vectors are given by\footnote{The vector quantities are formed with the $(x,\psi,y)$ components of the corresponding vector, for example $\vec{k} = (k^x,k^{\psi},k^y)$.}:
\begin{equation}
	\vec{k} = \begin{pmatrix} 0 \\ \alpha \\ 0 \end{pmatrix} + \begin{pmatrix} 0 \\ \beta \\ 0 \end{pmatrix} \times \begin{pmatrix} x \\ \psi \\ y \end{pmatrix},
\end{equation}
where the symbol $\times$ denotes the three-dimensional cross product.
In the $U(1)^3$ case, we need to specify how we gauge these isometries for each of the three vectors, so that we have a triplet of vectors $k^X_I$, $I=0,A,B$, and correspondingly $\alpha_I$ and $\beta_I$. A straightforward computation gives the corresponding moment maps:
\begin{equation}
	\vec{P}_I = \left( \beta_I + (2 \alpha_I - \beta_I\, r^2)  \frac{1}{\rho}\right)\begin{pmatrix} 0 \\ 1 \\ 0 \end{pmatrix} + \frac{1}{\sqrt{\rho}} \begin{pmatrix} 0 \\ \beta _I \\ 0 \end{pmatrix} \times \begin{pmatrix} x \\ \psi \\ y \end{pmatrix},
\end{equation}
where we have defined $r^2 = x^2 + y^2$.

In order to match the action (\ref{eqlagrPS}) with the on-shell two-derivative action (\ref{eq:actionRonshell}), we take:
\begin{align}
	g\alpha_3 &= \frac{m}{4}, & g\alpha_1 &=  -\frac{p_2}{8}, & g\alpha_2 &= -\frac{p_1}{8},\\
	g\beta_3 &= 0, & g\beta_1 &= m, & g\beta_2 &= m,
\end{align}
In fact, when $x = y = 0$, we have:
\begin{align}
\label{eq:su2PIs}	gP_3 & = m \frac{1}{\rho}, & gP_1 & = 2m -\frac{p_2}{2} \frac{1}{\rho}, & gP_2 &= 2m -\frac{p_1}{2}\frac{1}{\rho}.
\end{align}
Here we have defined $\vec{P}_I = P_I/2\,\begin{pmatrix} 1 \\ 0 \\ 0 \end{pmatrix}$, which leads to $P_{Iij}=P_I\delta_{ij}/2$. Moreover, we identify:
\be \rho = H^2.\ee
If we plug this, together with the moment maps \eqref{eq:su2PIs} and the metric \eqref{eq:su21u2metric}, into the Lagrangian \eqref{eq:actionR}, the resulting scalar potential does not depend on $\psi$. Furthermore, $x=y=0$ is an extremum of this potential.\footnote{We have checked that it is in fact a minimum of the potential for the solutions of interest in this paper.} It is now straightforward to check that for $x=y=0$ the resulting Lagrangian reproduces precisely the Lagrangian in \eqref{eqlagrPS}.	 

We note that while the above can be taken as evidence that the hyperscalars parametrize the manifold $SU(2,1)$/$U(2)$, it does not constitute a proof of this fact, since we have used a truncation with only one hyperscalar rather than the four required to form a complete hypermultiplet.

\subsection{The Chern--Simons terms at four derivatives}
In this section we compute the Chern--Simons couplings that appear in the 5d effective action by reducing the relevant terms directly from 11d. We follow closely the notation and results of \cite{Harvey:1998bx}. The main result that we use is the CS terms in 7d at leading and subleading order:
\begin{align}
	\label{eqn:CSaction}
	\mathcal{L}^{7d}_{(0)CS}+\mathcal{L}^{7d}_{(1)CS} = & \frac{2\pi N^3}{24} \left(\frac{m}{2}\right)^4 p_2^{(0)}(A) \nonumber\\
	& -  \frac{2\pi N}{48} \left(\frac{m}{2}\right)^4 \left(\left(\frac{2}{m}\right)^2 p_2(R) + p_2(A) - \left(\frac{2}{m}\right)^2\frac{p_1^2(R)}{4}\right.\\\nonumber
& \left. - \frac{p_1^2(A)}{4}+\left(\frac{2}{m}\right)\frac{p_1(R)p_1(A)}{2}\right)^{(0)},
\end{align}
where $p(A)$ and $p(R)$ are the Pontryagin classes built out of the $SO(5)$ and tangent bundle connections, respectively. 
The superscript $(0)$ simply means that we have to take the Chern--Simons form of the various terms, so that for example $d p_2^{(0)}(A) = p_2(A)$.

We only need:
\begin{align}
	p_1(A) &= -\frac12 \left(\frac{1}{2\pi}\right)^2 \mathrm{tr} F^2,\\
	p_2(A) &= \frac18 \left(\frac{1}{2\pi}\right)^4 \left((\mathrm{tr} F^2)^2-2 \mathrm{tr} F^4\right),
\end{align}
and analogous formulae for $p(R)$ where $F$ is replaced by $R$ (the curvature two-form). As a consequence, we have for example:
\begin{align}
	(p_1^2(A))^{(0)} & = \frac14 \left(\frac{1}{2\pi}\right)^4 \mathrm{tr}(A F)\mathrm{tr}(F^2)+\ldots,\\
	(p_2(A))^{(0)} & = \frac18 \left(\frac{1}{2\pi}\right)^4 \left(\mathrm{tr}(A F)\mathrm{tr}(F^2)-2 \mathrm{tr}(A F^3)\right)+\ldots,
\end{align}
where the ellipses denote terms that do not contribute when we truncate to the Cartan subalgebra of $SO(5)$.

We temporarily reintroduce the $SO(5)$ indices for the gauge fields: $A = A^{IJ}$. The two Cartan generators are taken to be:
\begin{align}
	\label{eq:normaliz}
	A^1 & \equiv \left(\frac{1}{2}\right) A^{12} & A^2 & \equiv \left(\frac{1}{2}\right) A^{34}.
\end{align}
The factor $1/2$ ensures that the normalization of the vector fields is consistent with the one we have been using for the 5d gauge fields, as we show at the end of this subsection. We have:
\begin{align}
	\mathrm{tr}(F^2) & = F^{IJ} F^{JI} = F^{12} F^{21} + F^{21} F^{12} + F^{34} F^{43} + F^{43} F^{34} + \ldots \\ & = - 8 (F^1)^2 - 8 (F^2)^2 + \ldots
\end{align}
where once again the ellipses denote generators that are not in the Cartan subalgebra. Analogously, we have:
\begin{align}
	\mathrm{tr}(AF) \mathrm{tr}(F^2) & = 4\times 16(A^1 F^1 + A^2 F^2)((F^1)^2 + (F^2)^2),\\
	\mathrm{tr}(A F^3) & =  2\times16 A^1 (F^1)^3 + 2\times16 A^2 (F^2)^3.
\end{align}

We can now write down the explicit form of the relevant CS terms:
\begin{align}
	\mathcal{L}^{7d}_{(0)CS} & =  \frac{N^3}{12\pi^3}\left(\frac{m}{2}\right)^4 \left(A^1 F^1 (F^2)^2 \right)\\
	\mathcal{L}^{7d}_{(1)CS} & = -\frac{N}{24\pi^3}\left(\frac{m}{2}\right)^4 \left(-\frac14 A^1 (F^1)^3 - \frac14 A^2 (F^2)^3 \right.\nonumber\\ & \qquad\qquad + \left.\frac12 A^1 F^1 (F^2)^2 - \frac{1}{4m^2} (A^1 F^1 + A^2 F^2) \mathrm{tr}(R^2) + \ldots\right),
\end{align}
where the ellipses denote terms that involve only the curvature two-form and which vanish when integrated over one Riemann surface.
It is now straightforward to reduce the above terms to 5d, using $F^A=p_A/2$ on the Riemann surface (which is consistent with (\ref{eq:7DreddefF})):
\begin{align}
\label{eq:5DCSanomaly0}
	\mathcal{L}^{5d}_{(0)CS} & =  \frac{N^3}{6 \pi^2}\left(\frac{m}{2}\right)^4\eta_2 \left(p_2 A^1 F^1 F^2 + p_1 A^1 F^2 F^2 \right),\\
\label{eq:5DCSanomaly1}
	\mathcal{L}^{5d}_{(1)CS} & = -\frac{N}{24 \pi^2}\left(\frac{m}{2}\right)^4\eta_2\left(- p_1 A^1 (F^1)^2 - p_2 A^2 (F^2)^2 + p_2 A^1 F^1 F^2 + p_1 A^1 F^2 F^2 \right. \nonumber\\ & \qquad\qquad \left. - \frac{1}{2m^2} (p_1 A^1 + p_2 A^2) \mathrm{tr}(R^2) + \ldots\right).
\end{align}
Here $\eta_2$ is related to the volume of the Riemann surface\footnote{We are working with a metric on $\Sigma$ of fixed scalar curvature $R_\Sigma=2\kappa$.  Then for $\mathfrak{g}\ne 1$, the formula for the Euler character gives:
\be
2-2\mathfrak{g}=\chi(\Sigma)=\frac{1}{4\pi}\int\sqrt{g_\Sigma}R_\Sigma=\frac{\kappa}{2\pi}\operatorname{vol}(\Sigma),
\ee
from which the result follows.  The convention for $\mathfrak{g}=1$ is fixed separately.} on which we have reduced from seven to five dimensions:
\be
\operatorname{vol}(\Sigma_2)=2\pi\eta_2=\left\{\begin{matrix} 2\pi, & \mathfrak{g}_2=1,\\ 4\pi\left|\mathfrak{g}_2-1\right|, & \mathfrak{g}_2\ne 1.\end{matrix}\right.
\ee
It is now straightforward to check that the normalization we have been using in \eqref{eq:normaliz} is the correct one. For the 7d system of M5 branes, we have:
\be G_N^{7d} = \left(\frac{2}{m}\right)^5 \frac{3\pi^2}{16} \frac{1}{N^3},\ee
which, using $G_N^{7d}=G_N^{5d}vol(\Sigma_2)$, leads to:
\be \label{eq:GN5} \frac{1}{16\pi G_N^{5d}} = \left(\frac{m}{2}\right)^5 N^3 \frac{2}{3\pi^2} \eta_2.\ee
Many conventions fix units as $m=2$, which is equivalent to fixing $R_{S^4}=1/2$ (and $R_{AdS_7}=1$ for the $AdS_7\times S^4$ solution) in the 11d reduction to 7d.
Our CS term coming from the two-derivative action $\mathcal{L}_R$ in \eqref{eq:actionR} is:
\begin{align} \mathcal{L}_R &\supset \frac{1}{2\kappa^2} C_{IJK} A^I\wedge F^J\wedge F^K\\
\label{eq:ourCSR} &= \frac{1}{16\pi G_N} \left(-\frac{k_1^{\psi}}{k_3^{\psi}} A^1 F^1 F^2 -\frac{k_2^{\psi}}{k_3^{\psi}} A^1 F^2 F^2\right),
\end{align}
where we have used the correct $C_{IJK}$ and used $k_I^{\psi}A^I=0$ to find the ``effective'' CS terms for $A^1,A^2$. Now, using \eqref{eq:GN5} and the expressions derived earlier for the constants $k_I^{\psi}$, we can see that our CS terms \eqref{eq:ourCSR} are identical to \eqref{eq:5DCSanomaly0}.

\subsection{Supersymmetric completion of the Chern--Simons terms at four derivatives}\label{sec:susycomplCS}
Once we have found all of the Chern-Simons terms at four-derivative order, in principle we can determine the full Lagrangian from supersymmetrizing these terms. This was originally done in $\mathcal{N}=1$ 5d supergravity for the four-derivative Chern-Simons term $A\wedge \Tr R^2$ in \cite{Hanaki:2006pj}, and recently more general four-derivative supersymmetric invariant Lagrangians in this theory have been constructed in \cite{Ozkan:2013nwa}. Note that the process of finding these completions crucially depends on the off-shell formalism, because the off-shell SUSY variations are independent of the action.

As explained in section 3, we only need the two supersymmetric higher-derivative Lagrangians found in \eqref{eq:actionC2} and \eqref{eq:actionR2}. The constants $c_I,b_I$ are a priori arbitrary constants, as they are not fixed by any SUSY considerations; in fact they depend on the details of the higher-dimensional theory that the 5d theory originates from. In our case, we can fix these constants by considering the coefficients of the Chern-Simons terms we found in the previous subsection. However, we are immediately faced with a  puzzle; equations \eqref{eq:actionC2} and \eqref{eq:actionR2} show that the higher-derivative corrections to the Chern--Simons terms are of the form (after integrating out the auxiliary field $V$):
\begin{align}
\label{eqn:4derCSstruct}
	d_I A^I \wedge P_J F^J \wedge P_K F^K,
\end{align}
where $d\in\{b,c\}$, and it is easy to see that there is no possible value for $d_I$ that can reproduce the terms in \eqref{eq:5DCSanomaly1}. 

However, we are overlooking another possible contribution to the order $N$ CS terms. Indeed, the Killing vector parameters $k_I^{\psi}$ are not determined by SUSY, but depend on the details of the compactification --- in our case they were determined from the reduction from 7d to 5d of the two-derivative 7d supergravity Lagrangian. This means that they too might receive corrections at higher order. We will allow $k_A^{\psi}$ (again, with $A\in\{1,2\}$) but not $k_3^{\psi}$ to receive corrections and parametrize them as:
\be k_A^{\psi} = k_A^{\psi(0)} + k_A^{\psi(1)}.\ee
This modifies the moment maps as:
\begin{align}
\label{eq:correctedPs}	gP_3 & = \frac{m}{H^2}, & gP_1 & = 2m -\frac{p_2}{2H^2} + k_1^{\psi(1)}\frac{4}{H^2}, & gP_2 &= 2m -\frac{p_1}{2H^2}+ k_2^{\psi(1)}\frac{4}{H^2}.
\end{align}
We need to clarify an important point. In the formula above we have used the relation between Killing vectors $k_I^X$ and moment maps $P_I$ induced by the leading order metric on the hyperscalar manifold. However, it is known that this metric receives higher-order corrections as well \cite{Strominger:1997eb,Gunther:1998sc,Antoniadis:2003sw,RoblesLlana:2006ez,Alexandrov:2007ec}. In our case we expect these corrections to shift the value of $H$ on-shell. Fortunately, as we will see in sections \ref{sec:4DcentralchargesSUGRA} and \ref{sec:2DcentralchargesSUGRA}, subleading corrections to $H$ do not affect the computation of the central charges.

The change in the Killing vectors will also modify the relation between $A^3$ and $A^A$ from the demand that $k^X_I A^I = 0$, giving:
\be A^3 = \frac{p_2}{2m}A^1 + \frac{p_1}{2m}A^2 -\frac{4k_1^{\psi(1)}}{m} A^1 -\frac{4k_2^{\psi(1)}}{m} A^2.\ee 
 In turn, this will contribute to the order $N$ CS terms when we integrate out $A^3$ through the two-derivative CS term $A^3 \wedge F^1 \wedge F^2$.
Fortunately, such corrections cannot give rise to terms like $A^A \wedge F^A \wedge F^A$, so the latter must necessarily come from the four-derivative CS terms in (\ref{eq:actionC2}) and (\ref{eq:actionR2}). In fact, we will now show that we can determine the parameters $c_I$, $b_I$ and $k_I^{\psi(1)}$ uniquely.

All of the CS terms in the our Lagrangian $\mathcal{L}_{tot}=\mathcal{L}_R + \mathcal{L}_{C^2}+\mathcal{L}_{R^2}$ in \eqref{eq:actionR}, \eqref{eq:actionC2}, and \eqref{eq:actionR2} are:
\begin{align}
\label{eq:ourCS1}\frac{1}{2\kappa_2}\mathcal{L}_{tot} &\supset \frac{1}{2\kappa^2}\left( C_{IJK}A^I F^J F^K - \frac14c_IA^I \Tr R^2 - \frac16 c_I A^I\mathcal{F}\mathcal{F} - b_I A^I \mathcal{F}\mathcal{F}\right)\\
\label{eq:ourCS2}&= \frac{1}{2\kappa^2}\left( A^1 F^2 F^3 - \frac14c_IA^I\Tr R^2 - (\frac{1}{24}c_I+\frac{1}{4}b_I)P_JP_K A^IF^JF^K\right)\\
 \nn &= \frac{1}{2\kappa^2}\left( -\frac{k_1^{\psi}}{k_3^{\psi}} A^1 F^1 F^2 -\frac{k_2^{\psi}}{k_3^{\psi}} A^1 F^2 F^2-\frac14 \tilde{c}_A A^A \Tr R^2\right.\\
& \left.  - 4m^2\left( \frac{\tilde{c}_1}{24} + \frac{\tilde{b}_1}{4}\right) A^1 F^1 F^1 - 4m^2\left(\frac{2\tilde{c}_1+\tilde{c}_2}{24} + \frac{2\tilde{b}_1+\tilde{b}_2}{4}\right) A^1 F^1 F^2\right.\non\\
\label{eq:ourCS3} & \left.  - 4m^2\left( \frac{\tilde{c}_2}{24} + \frac{\tilde{b}_2}{4}\right) A^2 F^2 F^2 - 4m^2\left(\frac{2\tilde{c}_2+\tilde{c}_1}{24} + \frac{2\tilde{b}_2+\tilde{b}_1}{4}\right) A^2 F^2 F^1\right),
\end{align}
where wedges are understood, and again we use $A\in\{1,2\}$. In going from \eqref{eq:ourCS1} to \eqref{eq:ourCS2} we used the correct $C_{IJK}$ and the leading order equation of motion for $V$ (\ref{eq:REOMV}). Then, to go from \eqref{eq:ourCS2} to \eqref{eq:ourCS3} we used the relation $k_I^XA^I=0$ to eliminate $A^3$ in favor of $A^A$, and introduced the notation:
\be \tilde{c}_A = c_A - \frac{k_A^{\psi}}{k_3^{\psi}}c_3, \qquad \tilde{b}_A = b_A - \frac{k_A^{\psi}}{k_3^{\psi}}b_3,\ee
and moreover we have used that $P_A - P_3 k_A^{\psi}/k_3^{\psi}=2m$ (at least at leading order).

As discussed before, the leading order expressions for $k_I^{\psi}$ and thus for $P_I$ are such that the CS term at leading order \eqref{eq:5DCSanomaly1} is correct. Now, to find the six higher order coefficients $\tilde{c}_A,\tilde{c}_B,k_A^{\psi(1)}$, we simply need to compare the coefficients of the terms in our expression \eqref{eq:ourCS3} to the known coefficients of the CS terms in \eqref{eq:5DCSanomaly1} and solve the resulting six non-degenerate linear equations. The result is:
\begin{align}
\label{eq:solcIbI}\tilde{c}_A &= -\frac{1}{N^2}\frac{p_A}{4m^3} & \tilde{b}_A &= \frac13 \tilde{c}_A,\\
\label{eq:solkI} k_1^{\psi(1)} &=  \frac{1}{N^2}k_3^{\psi}\frac{p_1+p_2}{4m}=-\frac{1}{N^2}\frac{\kappa_2}{8m} & k_2^{\psi(1)} &= k_1^{\psi(1)}  .
\end{align}
Note that we have thus determined two out of three linear combinations of the $c_I$ (and $b_I$), so we have not completely determined these coefficients yet. However, we can easily see that the third linear combination of the $c_I$ (and $b_I$) will never play a role in calculating the on-shell actions (which will be necessary in the following to calculate the central charge): e.g. $c_I$ always enters the action (\ref{eq:actionC2}) contracted with $\rho^I$ or $A^I$ (or $Y^I$, but on the solutions we are interested in, $Y^I$ is proportional to one of these two by (\ref{eq:AdS5solY}) and (\ref{eq:AdS3solY})). Then, we see that e.g.:
\be c_I \rho^I = c_A \rho^A + c_3 \rho^3 = \tilde{c}_A\rho^A + \frac{c_3}{k_3^{\psi}}(k_I^{\psi}\rho^I) = \tilde{c}_A\rho^A,\ee
because $k_I^X\rho^I=0$ from SUSY for all the solutions we are considering. Thus, even though we have not fully determined $c_3$, its actual value is irrelevant to compute the central charges because $k_I^X\rho^I=k_I^XA^I_a=0$. Of course, the same reasoning applies to $b_3$.

\subsection{Four-dimensional central charges}\label{sec:4DcentralchargesSUGRA}

Calculating the central charges of 4d CFTs from the $AdS_5$ dual was originally discussed in \cite{Henningson:1998gx}, and further extended to include higher-derivative corrections in \cite{Blau:1999vz,Nojiri:1999mh,Fukuma:2001uf,Cremonini:2008tw}. We will use the notation of \cite{Cremonini:2008tw}. Here, we are interested in calculating the central charge for the 4d CFTs discussed in section \ref{sec:4DCFT}, which arise from $N$ M5-branes wrapping a Riemann surface with genus $\mathfrak{g}$ (and corresponding twist parameters $z,\kappa$). We will be using the explicit $AdS_5$ solutions we derived in section \ref{sec:AdS5sol}.

The effective gravitational Lagrangian can be written as:
\be e^{-1}\mathcal{L} = \frac{1}{16\pi G_N^{eff}}\left( R + 12g_{eff}^2  + \alpha R^2 +\beta R_{\mu\nu}^2+\gamma R_{\mu\nu\rho\sigma}^2\right).\ee
Our effective Lagrangian, which we obtain by integrating out everything except the gravitational parts, is given by:
\begin{align}
\nn e^{-1}\mathcal{L} &= \frac{1}{16\pi G_N}\left[ R\left(1-\frac{3}{32} b_P\right) + \left(3 P^2 +\frac{33}{64} b_P P^2\right) + R^2\left(\frac{9}{64} \frac{b_P}{P^2}+\frac{P}{48}\left[c_I(P_I)^{-1}\right]\right)\right.\\
&\left. - \left(\frac{P}{6}\left[c_I(P_I)^{-1}\right]\right)R_{\mu\nu}R^{\mu\nu}
 + \frac{P}{8}\left[c_I(P_I)^{-1}\right] R_{\mu\nu\rho\sigma}R^{\mu\nu\rho\sigma}\right].
\end{align}
Note that $\left[c_I(P_I)^{-1}\right]\equiv \sum_I c_I (P_I)^{-1}$, and we have used the quantities $P$ and $b_P$ as defined in (\ref{eq:defP}) and (\ref{eq:defbP}). To maintain the correct normalization for the Einstein-Hilbert term, our effective Newton constant is given by:
\be \label{eq:shiftGN} \frac{1}{G_N^{eff}} = \frac{1}{G_N}\left(1-\frac{3}{32} b_P\right).\ee
Note that this shift in $G_N$ affects the value of $g_{eff}$. We can then identify the $AdS$ radius $L$ from the relation:
\be g_{eff} = \frac{1}{L}\left(1 - \frac{1}{3 L^2}(10 \alpha + 2\beta + \gamma)\right),\ee
which gives:
\be L = \frac{2}{P} - \frac{b_P}{2 P},\ee
which coincides with the expression derived earlier for the $AdS_5$ radius in (\ref{eq:AdS5Lsol}).

Finally, we should have \cite{Cremonini:2008tw}:
\begin{align} a_{grav} &= \frac{\pi L^3}{8 G_N^{eff}}\left( 1 - \frac{4}{L^2}(10\alpha+2\beta+\gamma)\right) = \left(\frac{m}{2}\right)^5\left(\frac{4N^3}{3}\right)\eta_2 \left(\frac{2}{P}\right)^3\left(1-\frac94 b_P\right),\\
 c_{grav} &= \frac{\pi L^3}{8 G_N^{eff}}\left( 1 - \frac{4}{L^2}(10\alpha+2\beta-\gamma)\right) = \left(\frac{m}{2}\right)^5\left(\frac{4N^3}{3}\right)\eta_2 \left(\frac{2}{P}\right)^3\non\\
 & \times\left(1-\frac94 b_P + \frac{P^3}{4}\left[c_I(P_I)^{-1}\right]\right).
\end{align}

A final piece of information we need is an explicit expression for $H$, since it appears in the $P_I$'s given in (\ref{eq:correctedPs}). As mentioned in section \ref{sec:AdS5sol}, $H$ is found by solving the equation $k_I^{\psi}\rho^I(H)=0$, where we have put in the solutions (\ref{eq:AdS5rhosol}) for the $\rho^I$ in terms of $P_I$ (thus introducing $H$ through the $P_I$ to the equation). Then, the solution to $k_I^{\psi}\rho^I=0$ to leading order is:
\be \label{eq:solH5D} H^2 = \frac{1}{4m}\left( p_1 + p_2 + \sqrt{p_1^2-p_1p_2+p_2^2}\right).\ee
In principle, we could similarly determine the subleading expression for $H$ as well, but we omit its expression as the subleading piece of $H$ can be seen not to contribute to $P$ at subleading order.

We can now fill in the explicit expressions for $P_I$, taking into account the subleading corrections of (\ref{eq:correctedPs}) and using our expression for $H$ (\ref{eq:solH5D}), and the expressions (\ref{eq:solcIbI})-(\ref{eq:solkI}) for the coefficients $b_I, c_I,k_A^{\psi(1)}$. Doing this, and expressing all $p_A$'s in terms of $\kappa_2,z_2$ using (\ref{eq:defpi}), we find (leaving out the subscript $2$ for $\kappa,z$):
\begin{align}
\nn a_{grav} &=  \frac{N^3\eta}{96} \frac{3z^2(-3\kappa+\sqrt{3z^2+\kappa^2})+\kappa^2(\kappa+\sqrt{3z^2+\kappa^2})}{z^2}\\
& - \frac{N\eta}{32}\frac{ (\kappa+\sqrt{3z^2+\kappa^2})(z^2+\kappa^2)}{z^2}.\\
\nn c_{grav} &= \frac{N^3\eta}{96} \frac{3z^2(-3\kappa+\sqrt{3z^2+\kappa^2})+\kappa^2(\kappa+\sqrt{3z^2+\kappa^2})}{z^2}\\
& + \frac{N\eta}{96} \frac{z^2(\kappa-2\sqrt{3z^2+\kappa^2})-3\kappa^2(\kappa+\sqrt{3z^2+\kappa^2})}{z^2},
\end{align}
which gives us a perfect match to order $N$ with the field theory expressions for the central charges $a$ and $c$ for $\kappa=\pm 1$ given in (\ref{eq:FTa4DorderNkappa1})-(\ref{eq:FTc4DorderNkappa1}) and $\kappa=0$ given in (\ref{eq:FTa4DorderNkappa0})-(\ref{eq:FTc4DorderNkappa0}).

\subsection{Two-dimensional central charges}\label{sec:2DcentralchargesSUGRA}
Calculating central charges of 2d CFTs for Lagrangians with higher-derivative corrections (generalizing Brown-Henneaux \cite{Brown:1986nw}) was derived in \cite{Imbimbo:1999bj,Saida:1999ec,Kraus:2005vz} and reviewed in \cite{Kraus:2006wn}. We will use the notation of \cite{Castro:2007sd,Kraus:2006wn}. We will be calculating the central charges of the 2d CFTs discussed in section \ref{sec:2DCFT}, which arise from $N$ M5-branes wrapping two Riemann surfaces with genus $\mathfrak{g}_1,\mathfrak{g}_2$ (and corresponding parameters $z_1,\kappa_1,z_2,\kappa_2$). We will be using the explicit $AdS_3\times\Sigma_g$ solutions we derived in section \ref{sec:AdS3sol}; by convention, the $\Sigma_g$ of the 5d solution of section \ref{sec:AdS3sol} will be the \emph{first} Riemann surface with parameters $z_1,\kappa_1$ (meaning the reduction of the 7d system to 5d happened over the \emph{second} Riemann surface) --- of course, this does not affect the final answers at all.

The central charge can be calculated by giving the on-shell Lagrangian:
\be \label{eq:cRpcL} \frac12\left(c_R + c_L\right) =  -\frac{3}{8G_5} \left[e^{2g_0} Vol(\Sigma_1)\right] e^{3f_0} \mathcal{L}   = -8 N^3\left(\frac{m}{2}\right)^5 \eta_1 \eta_2 e^{2g_0}e^{3f_0}\mathcal{L}.\ee
The leading part of (\ref{eq:cRpcL}) will be given by the on-shell leading order Lagrangian $\mathcal{L}_R$ in (\ref{eq:actionR}). The subleading part will have three contributions: one from each of the higher order on-shell Lagrangians $\mathcal{L}_{C^2}$ in (\ref{eq:actionC2}) and $\mathcal{L}_{R^2}$ in (\ref{eq:actionR2}), and a third contribution from $\mathcal{L}_R$ coming from the subleading corrections to the $k_I^{\psi}$'s and corresponding changes to the $P_I$'s as given in (\ref{eq:correctedPs}).\footnote{Note that one might think that $(e^{2g_0}e^{3f_0}\mathcal{L}_R)$ actually receives more than only these subleading corrections, due to subleading corrections of the $\rho^I$'s, $e^{f_0}$, $e^{g_0}$, and $H$. However, it is quite easy to see that the (leading order) equations of motion actually imply that $\partial((e^{2g_0}e^{3f_0}\mathcal{L}_R))/\partial\rho^I = 0$ in our particular case (and similar for $e^{f_0}$, $e^{g_0}$ and $H$), so that these subleading corrections vanish. Thus, the only ones that survive are those due to the actual correction of the $k_I^{\psi}$'s and the corresponding changes to the $P_I$'s and $A^3$.}

We still need to specify the values of the parameters $a^I$, which are the fluxes through the Riemann surface in the $AdS_3$ solution given in section \ref{sec:AdS3sol}. As mentioned at the end of section \ref{sec:AdS3sol}, these parameters must satisfy two restrictions:
\begin{align}
 k_I^{\psi} a^I &= 0\\
\frac12 g a^I P_I &= -\kappa_1.
\end{align}
We take the first equation to determine $a^3$, so:
\be a^3 = -\frac{k_1^{\psi}}{k_3^{\psi}}a^1-\frac{k_2^{\psi}}{k_3^{\psi}}a^2.\ee
We can fill in this expression into the second equation to obtain:
\be -\kappa_1 = \frac12 g a^A \tilde{P}_A = m (a^1+a^2),\ee
where we have used the definition of $\tilde{P}$ as in section \ref{sec:susycomplCS}. Thus, we can parametrize $a^1,a^2$ as:
\be \label{eq:AdS3explicitaI} a^1 = -\frac{\kappa_1-z_1}{2m}, \qquad a^2 = -\frac{\kappa_1+z_1}{2m}.\ee
As an aside, note that this is perfectly consistent with (\ref{eq:7DreddefF}) and the fact that it should be equivalent to reduce (from 7d to 5d) over the first or the second Riemann surface.

Also, as in the above $AdS_5$ case, we still need to determine the explicit expression for $H$; as mentioned in section \ref{sec:AdS3sol}, $H$ is once again found by solving the equation $k_I^{\psi}\rho^I(H)=0$, where we have put in the solutions (\ref{eq:AdS3rhosol}) for the $\rho^I$ in terms of $P_I$ (thus introducing $H$ through the $P_I$ to the equation). Then, the solution to $k_I^{\psi}\rho^I=0$ to leading order is:
\be \label{eq:solH3D} H^2 = \frac{1}{2m}\frac{(a^1p_2)^2+(a^2p_1)^2+a^1a^2p_1p_2}{(a^2)^2p_1+(a^1)^2p_2}.\ee
In principle, we could similarly determine the subleading expression for $H$ as well, but we omit its expression as the subleading piece of $H$ can be seen not to contribute to the on-shell Lagrangian.

Now, we can evaluate (\ref{eq:cRpcL}) on-shell for our solution, filling in our expressions for $b_I,c_I,k_I^{\psi,(1)}$ that we found in (\ref{eq:solcIbI})-(\ref{eq:solkI}), as well as the explicit values for $a^I$ in (\ref{eq:AdS3explicitaI}) and the expression for $H$ found above in (\ref{eq:solH3D}). This gives us:
\begin{align}
\nn\frac12(c_R+c_L) &= \frac{N^3\eta_1\eta_2}{4} \frac{3z_1^2z_2^2+z_2^2\kappa_1^2-8z_1z_2\kappa_1\kappa_2+z_1^2\kappa_2^2+3\kappa_1^2\kappa_2^2}{\kappa_1\kappa_2-3z_1z_2}\\
\nn& + \frac{N\eta_1\eta_2}{8(\kappa_1\kappa_2-3z_1z_2)^2} \left(9z_1^3z_2^3+12z_1z_2^3\kappa_1^2+9z_1^2z_2^2\kappa_1\kappa_2-2z_2^2\kappa_1^3\kappa_2\right.\\
& \left.+12z_1^3z_2\kappa_2^2+3z_1z_2\kappa_1^2\kappa_2^2-2z_1^2\kappa_1\kappa_2^3-\kappa_1^3\kappa_2^3\right),
\end{align}
 which is once again a perfect match with the field theory expression (\ref{eq:FTcRpcL}).

Finally, the quantity $c_R-c_L$ is related to the gravitational anomaly of the field theory and thus should come from the coefficient of the gravitational CS term in 3d, which in turn should come from the (reduction of the) 5d mixed gauge-gravitational CS term $A\wedge \Tr R^2$. To integrate this CS term down to 3d, we need to integrate out the $A$ piece, which (after partial integration) will simply give a factor of $-a^I vol(\Sigma_1)$. Thus, the coefficient of the 3d gravitational CS term, which is proportional to $c_R-c_L$, is given by \cite{Kraus:2005zm,Solodukhin:2005ah,Solodukhin:2005ns,Kraus:2006wn}:
\begin{align}
\nn \frac{1}{96\pi}(c_L-c_R) =\frac{1}{16\pi G_5} \left(-\frac14\right) vol(\Sigma_1) (-c_Ia^I) &= \frac{N^3}{3\pi} \left(\frac{m}{2}\right)^5\eta_1 \eta_2\ c_I a^I \\
&= -\frac{N}{384\pi} \eta_1 \eta_2 (z_1z_2+\kappa_1\kappa_2),
\end{align}
which matches with the field theory result (\ref{eq:FTcLmcR}).

\section{Conclusions}
In this paper we have considered higher-derivative corrections to $\mathcal{N}=1$ supergravity in five dimensions. Using off-shell techniques, we have been able to compute the corrections to $AdS_5$ and $AdS_3 \times \Sigma_g$ geometries in the presence of gauged isometries as well as non-trivial hypermultiplets.

The main application of our results is for various supersymmetric setups involving M5-branes wrapped around one or two Riemann surfaces. We were able to reproduce the first subleading corrections to the central charges of the dual SCFTs, which are known exactly from $a$-maximization and $c$-extremization. We extracted the precise data needed to characterize the supergravity corrections from the subleading corrections to the Chern--Simons terms, which in turn can be derived from the CP-odd eight-derivative correction of eleven-dimensional supergravity. A very intriguing outcome of our analysis is that the Killing vectors associated to the gauged isometries also receive $1/N$ corrections. This is very reminiscent of the analogous corrections to the universal hypermultiplet metric which were analyzed in \cite{Strominger:1997eb,Gunther:1998sc,Antoniadis:2003sw,RoblesLlana:2006ez,Alexandrov:2007ec} in the context of Calabi--Yau compactifications of M-theory, where the changes to the quaternionic metric are proportional to the Euler characteristic of the compactification manifold. Whether this is just a coincidence or a sign of something deeper is a matter that we leave to future investigation.

There are many interesting questions left to explore. One is to understand what the gravity dual of $c$-extremization is. While the answer is known in the case of $a$-maximization \cite{Tachikawa:2005tq}, the analogous results for $c$-extremization \cite{Karndumri:2013iqa,Karndumri:2013dca} were analyzed only at the two-derivative level. In this paper we studied $O(N)$ corrections to the central charges, but it should also be possible to get a handle on the $O(1)$ corrections by employing techniques along the lines of \cite{Ardehali:2013gra,Ardehali:2013xya,Ardehali:2013xla}\footnote{We thank Phil Szepietowski for suggesting this to us.}. It would be intriguing to analyze the corrections to the Killing vectors in a more systematic way, and to try to understand their structure in more general gauged supergravity setups. Another very compelling direction would be to extend our results to asymptotically $AdS_5$ supersymmetric black holes, in analogy to what has been done for asymptotically flat black holes in ungauged supergravity \cite{Castro:2007hc}. In particular, it would be extremely important to determine whether these geometries remain supersymmetric when higher-derivative corrections are taken into account. If the answer turned out to be negative, this might constitute a first step towards resolving the $1/16$-BPS black hole puzzle in maximal five-dimensional supergravity \cite{Kinney:2005ej,Chang:2013fba}.
\begin{center}
\bf{Acknowledgements}
\end{center}
\medskip
We would like to thank A.~Castro, J.~de~Boer, D.~Butter, D.~Hofman, J.~Jottar, and C.~Peng for fruitful discussions. We are especially grateful to Phillip Szepietowski for valuable comments on the draft. MB is supported in part by a grant from the Swiss National Science Foundation. DR is supported by funding from the European Research Council, ERC grant agreement no.\ 268088-EMERGRAV.  BW is supported in part by the STFC Standard Grant ST/J000469/1 ``String Theory, Gauge Theory and Duality." This work is part of the research programme of the Foundation for Fundamental Research on Matter (FOM), which is part of the Netherlands Organisation for Scientific Research (NWO).

\appendix

\section{$SU(2)$ Conventions \& Variations}

\subsection{$SU(2)$ \& Spinor Conventions}
\label{app:SU2Conventions}
We use the same $SU(2)$ index conventions as \cite{Bergshoeff:2004kh}, see also \cite{Bergshoeff:2001hc} for even more details.

Indices $i,j,k$ will always denote $SU(2)$ indices and run over $1,2$. Lowering and raising $SU(2)$ indices happens with the $\epsilon$ symbol in usual NW-SE contractions, e.g.:
\be A^i = \epsilon^{ij}A_j, \qquad A_i = A^j\epsilon_{ji}, \qquad \epsilon_{12}=-\epsilon_{21}=\epsilon^{12}=1. \ee
Note that $\epsilon_{jk}\epsilon^{ik}=\delta_j^{ i}$.

Often we will deal with $SU(2)$ doublets that have a pair of symmetric $SU(2)$ indices, e.g. $A^{ij}$. We will also use the equivalent three-vector notation $\vec{A}$. In general switching between doublet $ij$ and vector $r$ indices is accomplished by:
\be V^{ij} = i V^r \sigma^r_{ij},\ee
where the regular Pauli matrices are defined with indices as $(\sigma^r)_i^{\ j}$. Note that, using $\epsilon$ to raise/lower indices, we have e.g. $\sigma^2_{ij}=i\delta_{ij}$.

Our spinor conventions can also be found explicitly in appendix A of \cite{Bergshoeff:2001hc}. Here we list the most important facts. We define the charge conjugation operator as:
\be (\lambda^i)^C = \alpha^{-1}(\mathbf{C}\gamma_0)^{-1}\epsilon^{ij}(\lambda^j)^*,\ee
where $\mathbf{C}$ is the unitary charge conjugation matrix and $\alpha=\pm 1$ or $\alpha=\pm i$ depending on conventions for complex conjugation. Symplectic Majorana spinors, the minimal spinors in 5d, are spinors $\lambda^i$ where for $i=1,2$ the resulting spinor has four complex components; however, they satisfy $\lambda^C=\lambda$ and thus symplectic Majorana spinors in 5d have only 8 independent real components in total.

\subsection{$SU(2)$ Structure of SUSY variations}
In this section we derive some general properties regarding the $SU(2)$ structure of the fermion SUSY variations for our ansatze in section \ref{sec:susysols}.

\subsubsection{$SU(2)$ Fermions}
\label{app:su2vars}
We first consider the SUSY variations of the $SU(2)$ symplectic Majorana fermions, namely $\psi_\mu^i$, $\chi^i$ and $\lambda_I^i$. 
They are of the general form:
\begin{equation}
	\label{eq:generalvar}
	\delta \phi^i = \left(A \delta_j^{\phantom{j}i}+ B (i \vec{s}\cdot \vec{\sigma})_j^{\phantom{j}i} \right)\, \epsilon^j = 0,
\end{equation}
where $\vec{s}$ is a (real) unit vector and $A$ and $B$ depend on the fields and are real. This particular form is not well-suited for explicit computations, because it is difficult to implement projection conditions. As a consequence, we define the two projectors:
\begin{align}
	2(P_+)_n^{\phantom{n}m} &= \delta_n^{\phantom{n}m} + (\vec{s}\cdot \vec{\sigma})_n^{\phantom{n}m},\\
	2(P_-)_n^{\phantom{n}m}  &= \delta_n^{\phantom{n}m} -  (\vec{s}\cdot \vec{\sigma})_n^{\phantom{n}m},
\end{align}
and the two spinors:
\begin{align}
	\epsilon_+^m & = (P_+)_n^{\phantom{i}m} \epsilon^n,\\
	\epsilon_-^m & = (P_-)_n^{\phantom{i}m} \epsilon^n,
\end{align}
which satisfy $(\vec{s}\cdot \vec{\sigma})_j^{\phantom{j}i}\epsilon_{\pm}^j = \pm \epsilon_{\pm}^i$ 
The original spinor can be recovered as:
\begin{equation}
	\epsilon^i = \epsilon_+^i + \epsilon_-^i.
\end{equation}
Using \eqref{eq:generalvar}, it is straightforward to show that:
\begin{equation}
	(A\pm i\, B) \epsilon_{\pm}^m = 0.
\end{equation}

It is easy to check explicitly that the spinors $\epsilon_{+}^1$ and $\epsilon_{+}^2$ are proportional to each other, and similarly for $\epsilon_-^1$ and $\epsilon_-^2$.
To study supersymmetric solutions, one can now impose the projection conditions on, say, $\epsilon \equiv \epsilon_{\pm}^1$ and work with the simple equation
\begin{equation}
	(A + i \, B) \epsilon = 0.
\end{equation}
We can explain the equation above in a different way, which will be useful in the next subsection. Writing the SUSY variation as $\delta \phi^i = Q_{j}^{\phantom{j}i} \epsilon^j$, we notice that the projectors $P_{\pm}$ have been designed to commute with $Q$:
\begin{equation}
	0 = (P_\pm)_i^{\phantom{i}k} Q_{j}^{\phantom{j}i} \epsilon^j = Q_{i}^{\phantom{i}k} (P_\pm)_j^{\phantom{i}i} \epsilon^j =  Q_{i}^{\phantom{i}k} \epsilon_{\pm}^i = (A + i B) \delta_i^{\phantom{i}k}\epsilon_{\pm}^i.
\end{equation}

Our ``diagonal'' $SU(2)$ ansatz with $V'^{ij}=Y'^{Iij}=P'^{ij}_I=0$ means that:
\begin{equation}
	\vec{s} = \begin{pmatrix} 0 \\ 1 \\ 0\end{pmatrix},
\end{equation}
so that the $SU(2)$ structure is aligned with $\sigma^2$, which leads to
\begin{equation}
	\epsilon_{+} = \epsilon^1 + i \epsilon^2.
\end{equation}

\subsubsection{The hyperino}
\label{app:hyperinovar}
The discussion in the previous section does not immediately apply to the hyperino $\zeta^A$, which is a $USp(2n_H)$ symplectic spinor rather than $SU(2)$. The variation reads:
\begin{equation}
	\label{eq:zetaAvariationagain}
	\delta \zeta^A = \frac12 i \gamma^a(\partial_a q^X + g A^I_a k^X_I)f_X^{iA}\epsilon_i+\frac12 g \rho^I k_I^X f_{iX}^A\epsilon^i.
\end{equation}
In order to work with $SU(2)$ structures, we use the vielbein:
\begin{equation}
	\label{eq:zetaYivariation}
	f^Y_{i A}\delta \zeta^A = \frac12\left( i \gamma^a(\partial_a q^X + g A^I_a k^X_I)-g \rho^I k_I^X\right) f^Y_{i A}f_{X}^{jA}\epsilon_j.
\end{equation}
This variation is of the form:
\begin{equation}
	f^Y_{i A}\delta \zeta^A =  T^X {Q_X^{\phantom{X}Y}}_i^{\phantom{i}j}\epsilon_j, \qquad T^X= \frac14 \left( i \gamma^a(\partial_a q^X + g A^I_a k^X_I)-g \rho^I k_I^X\right),
\end{equation}
where the $SU(2)$ matrix $Q$ is given by:
\begin{equation}
	{Q_X^{\phantom{X}Y}}_i^{\phantom{i}j} = 2 f^Y_{i A}f_{X}^{jA} = \delta_i^j \delta_X^Y +  {J_X^{\phantom{X}Y}}_i^{\phantom{i}j},
\end{equation}
where ${J_X^{\phantom{X}Y}}_i^{\phantom{i}j}$ are the complex structures of the physical hyperscalar manifold \cite{Bergshoeff:2004kh}.

If we want to make use of the projection conditions on the Killing spinor, the discussion in the previous subsection shows that we should demand that $T^X {Q_X^{\phantom{X}Y}}_i^{\phantom{i}j}$ commutes with the projectors $P_{\pm}$. The condition turns out to be:
\begin{equation}
	\label{eq:vecsvecJ0}
	\vec{s} \times \left(T^X \, \vec{J}_X^{\phantom{X}Y}\right)= 0.
\end{equation}
Using the fact that the $\vec{J}$'s obey the quaternionic algebra, one can prove that any vector $\vec{v}$ can be written as:
\begin{equation}
	\vec{v}\, \delta_X^{\phantom{X}Y} = \frac12 \vec{v} \times \vec{J}_X^{\phantom{X}Y} - \frac12 \vec{J}_X^{\phantom{X}Z} \times (\vec{v} \times \vec{J}_Z^{\phantom{Z}Y}).
\end{equation}
Using this relation, the Jacobi identity of the vector product, and \eqref{eq:vecsvecJ0}, one sees that the only solution to \eqref{eq:vecsvecJ0} is $\vec{s}=0$. As a consequence, the only (general) solution to \eqref{eq:zetaAvariationagain} is:
\begin{equation}
	\partial_a q^X + g A_a^I k^X_I = 0, \qquad k^X_I \rho^I = 0.
\end{equation}
With constant hyperscalars, these conditions reduce to (\ref{eq:kIrhoI0}) and (\ref{eq:kIAI0}).

\section{Details on $\mathcal{N}=1$ Superconformal Supergravity}
\label{app:confSUGRA}

We use the superconformal actions and variations from \cite{Bergshoeff:2004kh} (mainly appendices A \& B therein), following their notation with the main exceptions that $\sigma^I_{theirs}=\rho^I_{ours}$ and $\psi^{iI}_{theirs}=\lambda^{iI}_{ours}$.

\subsection{Superconformal Action and Variations}
We have a Weyl multiplet with fields $e^a_{\mu},\psi^i_{\mu},V_{\mu}^{ij},T_{ab},\chi^i,D,b_{\mu}$; $n_v+1$ $U(1)$ vector multiplets with fields $A_{\mu}^I,Y_{ij}^I,\lambda^{I,i}, \rho^I$; and $n_H+1$ hypermultiplets with fields $q^{\hat{X}}, \zeta^{\hat{A}}$ (the hatted indices go over the full $n_H+1$ hypermultiplets, while unhatted indices will only run over the $n_H$ physical ones). To make further contact with \cite{Bergshoeff:2004kh}, note that we are not considering any tensor multiplets; moreover, because we have $U(1)$ gauge fields, $t_{IJ}^{\ K} = f_{IJ}^{\ K} = 0$.

One can construct a supersymmetric Lagrangian for the vector multiplets given a symmetric tensor $C_{IJK}$, and we define $\mathcal{C} = C_{IJK} \rho^I \rho^J \rho^K$ and $\mathcal{C}_I = 3 C_{IJK} \rho^J\rho^K, \mathcal{C}_{IJ} = 6C_{IJK} \rho^K$. The superconformal vector multiplet bosonic action is then given by \cite{Bergshoeff:2004kh}:\footnote{We will already put $b_{\mu}=0$ in some relations in anticipation of the superconformal gauge fixing. However, note that $\mathcal{D}_ab_{\mu}\neq0$ even if $b_{\mu}=0$. See \cite{Bergshoeff:2004kh} for more details.}
\begin{align}
\nn \mathcal{L}_{vector} &= \frac14 \mathcal{C}_{IJ} F^I\cdot F^J +\frac12\mathcal{C}_{IJ} \partial_a\rho^I\partial^a\rho^J - \mathcal{C}_{IJ} Y_{ij}^I Y^{ij J} + 8\mathcal{C}(D + \frac{26}{3} T^2+\frac{1}{32}R)\\
& - 8 \mathcal{C}_K F^K\cdot T +\frac{1}{4} e^{-1} \epsilon^{\mu\nu\rho\sigma\tau} C_{IJK} A^I_{\mu}F^J_{\nu\lambda}F^K_{\rho\sigma}.
\end{align}

For the hypermultiplets, there are a number of quantities that are relevant. First, there is a hypermultiplet metric $g_{\hX\hY}$ with corresponding vielbeins $f_{\hX}^{i\hA}$; this should define a hyperk\"ahler manifold \cite{Bergshoeff:2004kh,deWit:2001dj}. There are generators of dilatation and $SU(2)$ symmetries given by resp. $k^{\hX}, k_{ij}^{\hX}$. There are also Killing vectors $k_I^{\hX}$, which describe how the hyperscalars are charged under the vector multiplet gauge group; these (together with the complex structures of the hyperscalar manifold) also determine the moment maps $P_I^{ij}$.  Note that $\hX$ indices are raised or lowered with the metric $g_{\hX\hY}$, so e.g. $k^2 = k^{\hX}k_{\hX} = g_{\hX\hY}k^{\hX}k^{\hY}$. For more information on these hyperscalar quantities, see (especially sections 2.3.2-2.3.3 and section 3.3.2 of) \cite{Bergshoeff:2002qk}. The superconformal hypermultiplet bosonic action is then given by \cite{Bergshoeff:2004kh}:
\begin{align}
\nn \mathcal{L}_{hyper} &= -\frac12 g_{\hX\hY} (\partial_a q^{\hX} -V^{jk}_a k_{jk}^{\hX} + gA_a^I k_I^{\hX})(\partial^a q^{\hY} -V^{a\, jk} k_{jk}^{\hY} + gA^{Ia} k_I^{\hY}) + \frac49 D k^2\\
& + \frac{8}{27} T^2 k^2 - \frac{1}{24} R k^2 + 2g Y^{ij}_I P^I_{ij} - \frac12 g^2 \rho^I \rho^J k_I^{\hX} k_{J\ \hX}.
\end{align}

The relevant total superconformal action before gauge-fixing will be given by:
\begin{align}
\mathcal{L}_{SC,total} &= \mathcal{L}_{vector} + \mathcal{L}_{hyper}
\end{align}

The multiplets we are using are superconformal multiplets, which means they transform under with regular supersymmetries $Q$ with parameters $\epsilon^i$ as well as superconformal symmetries $S$ with parameters $\eta^i$. The superconformal fermionic variations on a bosonic background are given by \cite{Bergshoeff:2004kh}:
\begin{align}
\delta \psi^i_{\mu} &= (\partial_{\mu} + \frac14 \omega_{\mu}^{ab}\gamma_{ab})\epsilon^i - V^{ij}_{\mu}\epsilon_j + i \gamma\cdot T \gamma_{\mu} \epsilon^i - i\gamma_{\mu}\eta^i,\\
\nn \delta\chi^i &= \frac14 \epsilon^i D - \frac{1}{64} \gamma\cdot \mathcal{F}^{ij}\epsilon_j + \frac18 i \gamma^{ab} \slashed{\nabla}T_{ab}\epsilon^i - \frac18 i \gamma^a \nabla^b T_{ab} \epsilon^i\\
& - \frac14 \gamma^{abcd} T_{ab} T_{cd} \epsilon^i + \frac16 T^2 \epsilon^i + \frac14 \gamma\cdot T \eta^i,\\
\delta\lambda^{iI} &= -\frac14 \gamma\cdot F^I\epsilon^i - \frac12 i \slashed{\partial}\rho^I\epsilon^i-Y^{ij I}\epsilon_j + \rho^I \gamma\cdot T \epsilon^i + \rho^I\eta^i,\\
\delta \zeta^{\hA} &= \frac12 i \gamma^a(\partial_a q^{\hX}-V^{jk}_a k_{jk}^{\hX} + g A^I_a k^{\hX}_I)f_{\hX}^{i\hA}\epsilon_i -\frac13 \gamma\cdot T k^{\hX} f_{i\hX}^{\hA}\epsilon^i + \frac12 g \rho^I k_I^{\hX} f_{i\hX}^{\hA}\epsilon^i + k^{\hX} f_{i\hX}^{\hA}\eta^i.
\end{align}

\subsection{Gauge-fixing to Poincar\'e Supergravity}
\label{app:gaugefix}
To go from superconformal supergravity to the regular Poincar\'e supergravity, we need to gauge fix the (super)conformal symmetries. This is done by identifying one of the the hypermultiplets as non-physical and fixing it in order to fix the superconformal symmetries. This procedure is a bit involved; we sketch the highlights of it here but refer to \cite{Bergshoeff:2004kh} (especially section 4) for more details and derivations regarding this gauge-fixing procedure. See also appendix \ref{app:hypers} on more information regarding the hyperscalar manifold and the gauge fixing.

\subsubsection*{Splitting of hypermultiplets}
We split the hypermultiplet coordinates into $\hX = (x,X)$ where $x = 1,\cdots,4$ and $X=5,\cdots, 4(n_H+1)$. The hyperscalars are then given by:
\be q^{\hX} = (z^0, z^{\alpha}, q^X).\ee
The metric $g_{\hX\hY}$ splits as (see appendix \ref{sec:apphypersymmetries}):
\begin{align}
\nn d\hat{s}^2 &= g_{\hX\hY}dq^{\hX}dq^{\hY}\\
&=  - \frac{(dz^0)^2}{z^0} + z^0\left\{ h_{XY} dq^X dq^Y -g_{\alpha\beta}\left[ dz^{\alpha} + A^{\alpha}_X dq^X\right]\left[dz^{\beta} + A^{\beta}_Y dq^Y\right]\right\},
\end{align}
which essentially defines the $SU(2)$ connections $A^{\alpha}_X$ and the metrics $g_{\alpha\beta}, h_{XY}$. The $SU(2)$ connection $\omega^{ij})_X$ is given by:
\be \vec{\omega}_X = -\frac12 \vec{A}_X.\ee 
Note that it is the metric $h_{XY}$ on the physical hypermultiplet space that must be quaternionic \cite{Bergshoeff:2004kh,deWit:2001dj}; there are also corresponding complex structures ${J_X^{\phantom{X}Y}}_i^{\phantom{i}j}$ that satisfy the quaternionic algebra (see also appendix \ref{app:hypers}). The explicit vielbeins of the hyperscalar metric are needed for computations are are listed in section 3 of \cite{Bergshoeff:2004kh}.

 The quantities $k^{\hX}$ and $k^{\hX}_{ij}$ are the dilatation and $SU(2)$ transformations, and we can choose them to be:
\be k^{\hX} = (3z^0, 0, 0), \qquad k^{\hX}_{ij} = (0,k^{\alpha}_{ij},0).\ee
The Killing vectors $k_I^{\hX}$ split as:
\be k_I^{\hX} = (0, -2\vec{k}^{\alpha}\cdot (\vec{\omega}_Xk_I^X-\frac{1}{z^0}\vec{P}_I),k_I^X).\ee
Note that e.g.:
\begin{align} k^2 &= g_{\hX\hY} k^{\hX} k^{\hY} = -9z^0,\\
k^{\hat{X}}_I k_{\hat{X},J} &= 2h_{XY}k^X_I k^Y_J + (\vec{k}^{\alpha}\cdot\vec{P}_I)(\vec{k}_{\alpha}\cdot\vec{P}_J).
\end{align}

The fermionic sector is split into:
\be \zeta^{\hA} = (\zeta^i, \zeta^A),\ee
where $i=1,2$ and $A=1, \cdots,2n_H$. Only $\zeta^i$ should transform under superconformal $S$-transformations.

\subsubsection*{$K$-gauge}
This is fixed by setting $b_{\mu}=0$. Keeping this gauge fixed, i.e. $\delta b_{\mu}=0$, fixes the superconformal transformation parameter $\Lambda_{K\mu}$ which we have ignored in the previous discussion as it is not relevant for us.

\subsubsection*{$D$-gauge}
To get a factor of $1$ multiplying $R$ in the two-derivative action (on-shell, after imposing the equation of motion for $D$ which will be $\mathcal{C}=1$), we want to set $k^2 = -18$, i.e.:
\be z^0 = 2 .\ee

\subsubsection*{$SU(2)$-gauge}
We fix:
\be q^{\alpha}_{ij} = z^{\alpha}_{ij} = z^{\alpha}_{ij,0},\ee
i.e., they are constants.

The vector $\vec{k}^{\alpha}$ generates an $SU(2)$ algebra and are left-invariant vector fields. We can then choose the constants $z^{\alpha}_{ij,0}$ such that (see appendix \ref{sec:embedding}):
\be \label{eq:gaugechoice} \vec{k}^{\alpha} = k^{r,\alpha} = \delta^{r\alpha},\ee
to make expressions involving $\vec{k}^{\alpha}$ simple. This also means:
\be k^{\alpha}_{ij} = i \sigma^{\alpha}_{ij}.\ee

\subsubsection*{$S$-gauge}
Finally, we want to fix:
\be \zeta^i = 0,\ee

Keeping the $S$-gauge fixed means fixing $\delta \zeta^i = 0$, which will fix the $S$-transformation parameter $\eta^i$ as a function of $\epsilon^i$:
 \begin{align}
\nn \delta \zeta^k &= -\frac14 i  \gamma^a\left[ k^{\alpha}_{lm} V^{lm}_a - \mathcal{A}_X^{\alpha} \partial_a q^X - g A^I_a (\vec{k}^{\alpha}\cdot \vec{P}_I)\right](\vec{k}_{\alpha}\cdot\vec{\sigma}^{jk})\epsilon_j - i\gamma\cdot T\epsilon^k + 3i \eta^k\\
& - \frac14 g \rho^I (\vec{k}^{\alpha}\cdot\vec{P}_I)(\vec{k}_{\alpha}\cdot\vec{\sigma}^{jk})\epsilon_j.
\end{align}
Setting this to zero has as solution:
\begin{align}
 \eta^i &= \frac{i}{6}  \gamma^a\Upsilon_a^{ji}\epsilon_j +\frac13\gamma\cdot T\epsilon^i  + \frac{1}{6} g \rho^I P_I^{ji}\epsilon_j.
 \end{align}

\subsubsection*{Gauge-fixed Lagrangian \& Variations}
Using the above formulae to fix the relevant quantities, we find the gauge-fixed Poincar\'e supergravity Lagrangian given by (\ref{eq:actionR}). Also, fixing $\eta^i$ as given above gives us the gauge-fixed supersymmetry variations given by (\ref{eq:vargravitino})-(\ref{eq:varhyperino}).

\subsection{Full Two-derivative Equations of Motion}
\label{app:twoderEOM}
The full two-derivative equations of motion that follow the action (\ref{eq:actionR}) are:
\begin{align}
0 &= 8\lp\mathcal{C}-1\rp,\non\\
0 &= 4\Upsilon_a^{ij},\non\\
0 &= -2\mathcal{C}_{IJ}Y^{J\,ij}+2gP_I^{ij},\non\\
0 &= \frac{32}{3}\lp 13\mathcal{C}-1\rp T_{ab}-8\mathcal{C}_IF^I_{ab},\non\\
0 &= \mathcal{C}_{IJ}\nabla_bF^{J\,ab}+6C_{IJK}\nabla_b\rho^JF^{K\,ab}-16\mathcal{C}_I\nabla_bT^{ab}-16\mathcal{C}_{IJ}\nabla_b\rho^JT^{ab}\non\\
& +\frac{3}{4}\e^{abcde}C_{IJK}F^J_{bc}F^K_{de}-2gh_{XY}k_I^X\lp\p^aq^Y+gA^{J\,a}k_J^Y\rp-2gP_I^{ij}\Upsilon^a_{ij},\non\\
\label{eq:appEOMD}0 &= \frac{3}{2}C_{IJK}F^{J\,ab}F^K_{ab}-3C_{IJK}\p^a\rho^J\p_a\rho^K-\mathcal{C}_{IJ}\nabla^2\rho^J-6C_{IJK}Y^{J\,ij}Y^K_{ij}\\
& +\mathcal{C}_I\lp 8D+\frac{208}{3}T^{ab}T_{ab}+\frac{1}{4}R\rp-8\mathcal{C}_{IJ}F^J_{ab}T^{ab}-2g^2\rho^Jk_I^Xk_J^Yh_{XY}+g^2\rho^JP_I^{ij}P_{J\,ij},\non\\
0 &= \hlf g^{\m\n}\ls\frac{1}{4}\mathcal{C}_{IJ}F^{I\,ab}F^J_{ab}-\mathcal{C}_{IJ}Y^{I\,ij}Y^J_{ij}+8\lp\mathcal{C}-1\rp D+\frac{16}{3}\lp 13\mathcal{C}-1\rp T^{ab}T_{ab}\right.\non\\
& \left.+\frac{1}{4}\lp\mathcal{C}+3\rp R-8\mathcal{C}_IF^I_{ab}T^{ab}-h_{XY}\lp\p^aq^X+gA^{I\,a}k_I^X\rp\lp\p_aq^Y+gA_a^Jk_J^Y\rp\right.\non\\
& \left.+2gY^I_{ij}P_I^{ij}-g^2\rho^I\rho^Jk_I^Xk_J^Yh_{XY}+\hlf g^2\rho^I\rho^JP_I^{ij}P_{J\,ij}+2\Upsilon^a_{ij}\Upsilon^a_{ij}-\hlf\mathcal{C}_I\nabla^2\rho^I\rs\non\\
& -\frac{1}{4}\lp\mathcal{C}+3\rp R^{\m\n}+\frac{1}{4}\mathcal{C}_I\nabla^\m\nabla^\n\rho^I-\frac{1}{4}\mathcal{C}_{IJ}\p^a\rho^I\p_a\rho^J-\hlf\mathcal{C}_{IJ}F^{I\,\m\rho}F^{J\,\n}_{\hph{J\,\n}\rho}\non\\
& -\frac{32}{3}\lp 13\mathcal{C}-1\rp T^{\m\rho}T^\n_{\hph{\n}\rho}+16\mathcal{C}_IF^{I\,(\m|\rho|}T^{\n)}_{\hph{\n)}\rho}\non\\
& +h_{XY}\lp\p^\m q^X+gA^{I\,\m}k_I^X\rp\lp\p^\n q^Y+A^{J\,\n}k_J^Y\rp-2\Upsilon^{\m\,ij}\Upsilon^\n_{ij},\non
\end{align}
along with the equation of motion for the hyperscalars, which we omit. We can solve (\ref{eq:appEOMD}) for $D$:
\begin{align}
D &= \frac{1}{284}\lp 2\mathcal{C}_I\mathcal{C}_J-3\mathcal{C}_{IJ}\rp F^{I\,ab}F^J_{ab}+\frac{g^2}{64}\lp 6\lp\mathcal{C}^{-1}\rp^{IJ}-\rho^I\rho^J\rp P_I^{ij}P_{J\,ij}+\frac{g^2}{32}\rho^I\rho^Jk_I^Xk_J^Yh_{XY}\non\\
 & -\frac{3}{64}\mathcal{C}_{IJ}\p^a\rho^I\p_a\rho^J-\frac{1}{32}h_{XY}\lp\p^aq^X+gA^{I\,a}k_I^X\rp\lp\p_aq^Y+gA_a^Jk_J^Y\rp,
\end{align}

\section{7d Supergravity Conventions}\label{app:7Dgauged}
Here we give a quick sketch of the 7d $U(1)^2$ gauged supergravity theory that we are considering. This theory is a truncation of $SO(5)$ gauged maximal supergravity in 7d, so we first give an overview of this theory. We give the relevant references where more information on these theories can be found.

\subsection{Gauged Maximal Supergravity}
The theory that is obtained by reducing M-theory on an $S^4$ is gauged $\mathcal{N}=4$ (maximal) supergravity in 7d. This was first derived in \cite{Pernici:1984xx}; we will sketch the (bosonic) theory here using the notation of \cite{Liu:1999ai}.

The Lagrangian takes the form \cite{Pernici:1984xx, Liu:1999ai}:
\begin{align}
\nn 2\kappa^2 e^{-1}\mathcal{L}_{7d} &= R + \frac12 m^2 (T^2-2T_{ij}T^{ij}) - \Tr (P_{\mu}P^{\mu}) - \frac12 (V_I^{\ i} V_J^{\ j} F_{\mu\nu}^{IJ})^2 + m^2 (V_i^{-1\ I} C^I_{\mu\nu\rho})^2\\
 & + e^{-1}\left(\frac12\delta^{IJ} (C_3)_I\wedge (dC_3)_J + m\ \epsilon_{IJKLM} (C_3)_I F^{JK} F^{LM} + m^{-1}p_2(A,F)\right).
\end{align}
The gauge group is $SO(5)_g$ and $I, J\in \{1,\cdots, 5\}$ are fundamental indices of this group. The bosonic field content consists of the graviton, ten Yang-Mills gauge fields $A^{IJ}$ in the adjoint of $SO(5)_g$, five antisymmetric three-tensors $(C_3)_I$ in the fundamental of $SO(5)_g$, and 14 scalars which parametrize a $SL(5,\mathbb{R})/SO(5)_c$ coset; $i,j=\{1,\cdots, 5\}$ are fundamental indices of $SO(5)_c$. These scalar degrees of freedom are contained in $V_I^{\ i}$, an element in the coset; the other relevant scalar quantities are defined through:
\be V_i^{-1\ I}\mathcal{D}_{\mu}V_I^{\ j} = (Q_{\mu})_{[ij]} + (P_{\mu})_{(ij)},\ee
where $\mathcal{D}$ is the fully gauge-covariant derivative so that $\mathcal{D}_{\mu}V_I^{\ i} = \partial_{\mu}V_I^{\ j} + (2m)A_{\mu\ I}^J V_J^{\ j}$. The $T$-tensor is defined as:
\be T_{ij} = V_i^{-1\ I} V_j^{-1\ J}\delta_{IJ}, \qquad T = T_{ij}\delta^{ij}.\ee
The gauge coupling $m$ is related to the radius of the $S^4$ in 11d by $R_{S^4}=1/m$ \cite{Cvetic:1999xp}. Finally, $p_2(A,F)$ denotes the CS terms involving only the gauge fields $F^{IJ}_{\mu\nu}$; these are discussed in detail in section 4.2.

The fermionic variations are given by:
\begin{align}
 \nn \delta\psi_{\mu} &= \left[ \mathcal{D}_{\mu} + \frac{m}{20} T\tilde{\gamma}_{\mu} - \frac{1}{40}(\tilde{\gamma}_{\mu}^{\ \nu\lambda} - 8\delta^{\nu}_{\mu}\tilde{\gamma}^{\lambda})\Gamma^{ij}V_I^{\ i}V_J^{\ i} F_{\nu\lambda}^{IJ}\right.\\
 &\left. + \frac{m}{10\sqrt{3}}(\tilde{\gamma}_{\mu}^{\ \nu\lambda\sigma}-\frac92 \delta^{\nu}_{\mu}\tilde{\gamma}^{\lambda\sigma})\Gamma^iV_i^{-1\ I} C_{\nu\lambda\sigma}^I\right]\epsilon,\\
\nn \delta\lambda_i &= \left[ \frac{m}{2}(T_{ij}-\frac15\delta_{ij}T)\Gamma^j + \frac12 \tilde{\gamma}^{\mu}P_{\mu ij}\Gamma^j + \frac{1}{16}\tilde{\gamma}^{\mu\nu}(\Gamma^{kl}\Gamma^i-\frac15\Gamma^i\Gamma^{kl})V_K^{\ k}V_L^{\ l}F_{\mu\nu}^{KL} \right.\\
 &\left. +\frac{m}{20\sqrt{3}}\tilde{\gamma}^{\mu\nu\lambda}(\Gamma^{ij}-4\delta^{ij})V_j^{-1\ J} C^J_{\mu\nu\lambda}\right]\epsilon,
\end{align}
for resp.\ the gravitini and gaugini. The supersymmetry parameter $\epsilon$ transforms as the spinor of $SO(5)_c$ with corresponding Dirac matrices $\Gamma^i$; $\tilde{\gamma}^{\mu}$ are the 7d spacetime gamma matrices.

\subsection{$U(1)^2$ Truncation}
There is a simple truncation of the field content to a $U(1)^2$ gauge group, as described in \cite{Maldacena:2000mw,Liu:1999ai} and also used in \cite{Szepietowski:2012tb}. We will sketch the relevant information about this truncation from \cite{Liu:1999ai} here.

We restrict ourselves to the Cartan gauge fields:
\be A^{(1)}_{\mu} \equiv A^{12}_{\mu}, \qquad A^{(2)}_{\mu} \equiv A^{34}_{\mu},\ee
and also restrict the scalars to:
\be V_I^{\ i} = \rm{diag}\left( e^{-\lambda_1}, e^{-\lambda_1}, e^{-\lambda_2}, e^{-\lambda_2}, e^{2\lambda_1+2\lambda_2}\right),\ee
thus defining the two independent scalars $\lambda_{(i)}$. We also restrict to a single three-form, $S_{\mu\nu\rho}\equiv C^5_{\mu\nu\rho}$.

Thus, this truncation contains two Abelian gauge fields, two scalars, and a single three-form. This does not necessarily correspond to a consistent truncation of the maximal theory, as discussed in \cite{Liu:1999ai}. The truncated bosonic Lagrangian has the form:
\begin{align}
\nn 2\kappa^2e^{-1} \mathcal{L} &= R -\frac12 m^2 V  - 5(\partial_{\mu}(\lambda_1+\lambda_2))^2 -(\partial_{\mu}(\lambda_1-\lambda_2))^2 - e^{-4\lambda_1}F_{(1)}^2 - e^{-4\lambda_2}F_{(2)}^2\\
\nn  & + m^2 e^{-4\lambda_1-4\lambda_2}S_{\mu\nu\lambda}^2 - \frac{m}{6}\epsilon^{\mu\nu\lambda\alpha\beta\gamma\delta}S_{\mu\nu\lambda}\partial_{\alpha}S_{\beta\gamma\delta}\\
& + \frac{1}{\sqrt{3}}\epsilon^{\mu\nu\lambda\alpha\beta\gamma\delta}S_{\mu\nu\lambda}F^{(1)}_{\alpha\beta}F^{(2)}_{\gamma\delta} + m^{-1}p_2(A,F),\\
V & = -8 e^{2\lambda_1+2\lambda_2} -4e^{-2\lambda_1-4\lambda_2} -4 e^{-4\lambda_1-2\lambda_2}+e^{-8\lambda_1-8\lambda_2}.
\end{align}
We will not give all of the explicit equations of motion following from this Lagrangian or the explicit expressions of the supersymmetry variations for this truncation; both can be found in \cite{Liu:1999ai}. Here we will only show the self-duality equation for the three-form (which is its equation of motion):
\be e^{-4\lambda_1-4\lambda_2}S = *\left(\frac{1}{m} dS  - \frac{2}{\sqrt{3}m^2} F^{(1)}\wedge F^{(2)}\right).\ee

\section{Hypercomplex and quaternionic geometries}\label{app:hypers}
In this appendix we collect some useful facts about the geometry of the hyperscalars. We will first describe hypercomplex and quaternionic geometries. We then describe how quaternionic manifolds can be embedded in hypercomplex manifolds with conformal symmetry. The precise map between the two allows us to show that the gauge-fixing can be chosen in a convenient way, thus justifying the choice made in \eqref{eq:gaugechoice}. We will not discuss various subtleties, such as $\xi$-transformations and the possibility that no metric exists, for the sake of clarity. We will use the notation and conventions of \cite{Bergshoeff:2004nf} throughout. For applications to supergravity, see \cite{VanProeyen:2001ng}, \cite{Bergshoeff:2002qk} and especially \cite{Bergshoeff:2004kh}.
\subsection{Quaternionic like manifolds}\label{app:hyperquaternionic}
On-shell local supersymmetry implies that the hyperscalars parametrize a quaternionic manifold. In the following, we will use local coordinates $q^X$, $X=1,\ldots,4r$, where $r$ is the number of hypermultiplets. Furthermore we always assume the existence of a (invertible) vielbein $f_X^{iA}$, $i=1,2$, $A = 1,\ldots,2r$. This quantity appears in the supersymmetry transformations of the hyperscalars:
\begin{equation}
	\delta q^X = f^{X}_{iA}\bar{\epsilon}^i \zeta^A,
\end{equation}
where the inverse vielbein is defined as:
\begin{equation}
	f_Y^{iA} f_{iA}^X = \delta_Y^X, \qquad f_X^{iA} f^X_{jB} = \delta_j^i \delta_B^A.
\end{equation}
Furthermore the vielbein satisfies a reality condition defined by the matrices $E_i^{\phantom{i}j}$ and $\rho_A^{\phantom{A}B}$ such that:
\begin{equation}
	E E^* = -\mathds{1}_2, \qquad \rho \rho^* = -\mathds{1}_{2r},
\end{equation}
namely:
\begin{equation}
	(f_X^{iA})^* = f_X^{jB}E_j^{\phantom{j}i}\rho_B^{\phantom{B}A}.
\end{equation}

The vielbein and its inverse can be used to define a quaternionic structure:
\begin{equation}
	\vec{J}_X^{\phantom{X}Y} \equiv - i f_X^{iA} \vec{\sigma}_i^{\phantom{i}j} f^Y_{jA},
\end{equation}
where $\vec{\sigma}$ are the Pauli matrices. The name ``quaternionic" comes from the fact that these quantities obey the quaternion algebra:
\begin{equation}
	{J^r}_X^{\phantom{X}Z} {J^s}_Z^{\phantom{Z}Y} = -\delta^{rs} {\delta_X}^Y + \epsilon^{rst} {J^t}_X^{\phantom{X}Y}.
\end{equation}
In order for the manifold to be \textit{quaternionic}, however, the quaternionic structure needs to be integrable, which amounts to the existence of a \emph{torsionless} affine connection ${\Gamma_{XY}}^Z$, a $G\ell(r,\mathbb{H})$ connection ${\omega_{X A}}^B$ and a $SU(2)$ connection ${\omega_{Xi}}^j$ with respect to which the vielbein is covariantly constant:
\begin{equation}
	\label{eq:vielbeincovconst}
	\mathfrak{D}_X f_Y^{iA} \equiv \partial_X f_Y^{iA} - {\Gamma_{XY}}^Z f_Z^{iA} + f_Y^{jA} {\omega_{Xj}}^i + f_Y^{iB}{\omega_{XB}}^A = 0.
\end{equation}
This equation tells us that the holonomy of the manifold is \emph{restricted}. Indeed, consider the spin connection defined as:
\begin{equation}
	{\Omega_{X jB}}^{iA} \equiv f^Y_{jB}(\partial_X f_Y^{iA} - {\Gamma_{XY}}^Z f_Z^{iA}).
\end{equation}
If ${\Omega_{X jB}}^{iA}$ were a general $4r \times 4r$ matrix, the holonomy would be (a generic subgroup of) $G\ell(4r)$. Equation \eqref{eq:vielbeincovconst} is equivalent to:
\begin{equation}
	{\Omega_{X jB}}^{iA} = -{\omega_{Xj}}^i {\delta_B}^A - {\omega_{XB}}^A {\delta_i}^j,
\end{equation}
where ${\omega_{Xj}}^i$ is traceless. This implies that the holonomy group is restricted to $SU(2)\times G\ell(r,\mathbb{H})$.
If the $SU(2)$ connection is zero (or rather, pure gauge), the manifold is called \textit{hypercomplex}.
The conditions above imply that the quaternionic structure is covariantly constant as well, in the sense that:
\begin{equation}
	\mathfrak{D}_X \vec{J}_Y^{\phantom{Y}Z} \equiv \partial_X \vec{J}_Y^{\phantom{Y}Z} - {\Gamma_{XY}}^W \vec{J}_W^{\phantom{W}Z} + {\Gamma_{XW}}^Z \vec{J}_Y^{\phantom{Y}W} + 2\, \vec{\omega} \times \vec{J}_Y^{\phantom{Y}Z} = 0.
\end{equation}

Quaternionic and hypercomplex manifolds that admit a Hermitian invertible metric $g$ compatible\footnote{The compatibility condition is quite subtle, see \cite{Bergshoeff:2004nf}, however we can roughly think of it as being the requirement that the affine connection in the formulae above is the Levi-Civita connection associated to this metric.} with the affine connection are called \textit{quaternionic-K\"ahler} and \textit{hyperk\"ahler} respectively. In an appropriate basis, such a metric can be written as
\begin{equation}
	\label{eq:metricquaternionic}
	g_{XY} = f_{X}^{iA} C_{AB} \epsilon_{ij} f_Y^{jB},
\end{equation}
where $C = \epsilon \otimes \mathds{1}_r$. In this case, the holonomy is further restricted to the maximal compact subgroup of $SU(2) \times G\ell(r,\mathbb{H})$, namely $SU(2) \times USp(2r)$. \footnote{In principle, one could have $USp(2p,2r-2p)$.}
Such manifolds are Einstein and the $SU(2)$ curvatures are proportional to the complex structures:
\begin{equation}
	R_{XY} = \frac{1}{4r} g_{XY} R, \qquad \vec{\mathcal{R}}_{XY}=\frac12 \nu \vec{J}_{XY}, \qquad \nu = \frac{1}{4r(r+2)} R.
\end{equation}
Notice also that in supergravity, supersymmetry connects $\nu$ to the normalization of the Einstein term in the action, so that we have $\nu = - \kappa^2$, $\kappa$ being the gravitational coupling constant.
\subsection{Conformal symmetry}
For the applications to the superconformal tensor calculus, we are interested in hypercomplex manifolds with conformal symmetry. We will see that it is always possible to embed any $4r$-dimensional quaternionic manifold (the \textit{small }space in the following) into a $(4r+4)$-dimensional hypercomplex manifold with conformal symmetry (the \textit{big} space in the following). We will denote quantities on the big space with hats; for example the coordinates on the small space will be denoted by $X=1,\ldots,4r$ and coordinates on the big space by $\hat{X}=1,\ldots,4r+4$.

Conformal symmetry is defined by the existence of a so called ``homothetic" Killing vector $k^X$ defined as:
\begin{equation}
	\mathfrak{D}_{\hat Y} k^{\hat X} \equiv \partial_{\hat Y}k^{\hat X} + {\Gamma_{\hat{Y}\hat{Z}}}^{\hat{X}} k^{\hat Z} = \frac32 {\delta_{\hat Y}}^{\hat X} .
\end{equation}
Three more vectors can be constructed:
\begin{equation}
	\vec{k}^{\hat X} \equiv \frac13 k^{\hat Y}\hat{\vec{J}}_{\hat{Y}}^{\phantom{Y}\hat{X}}.
\end{equation}
In the absence of a metric, one needs to impose some additional requirements that will not be important for us; see \cite{Bergshoeff:2004nf}.

One can choose coordinates so that the $k$'s take a convenient form. Concretely, we choose:
\begin{equation}
	q^{\hat X} = (z^0, z^\alpha, q^X), \qquad \alpha=1,2,3, \quad X=1,\ldots,4r,
\end{equation}
such that:
\begin{equation}
	k^{\hat X} = 3 z^0 \delta_0^{\hat X}, \qquad  \vec{k}^0 = \vec{k}^X = 0.
\end{equation}
Therefore the only non-zero components of the $\vec{k}$'s are $\vec{k}^\alpha$. One also introduces the inverse vectors $\vec{m}_\beta$ so that:
\begin{equation}
	\vec{k}^\alpha \cdot \vec{m}_\beta = \delta^\alpha_\beta.
\end{equation}
It is also convenient to define a vector $\vec{A}_X$ as:
\begin{equation}
	\vec{A}_X \equiv \frac{1}{z^0} \hat{\vec{J}}_X^{\phantom{X}0}
\end{equation}
This leads to the following decomposition for the quaternionic structure:
\begin{align}
\label{eq:hattedquaternionicstructure}
	\hat{\vec{J}}_{0}^{\phantom{0}0} & = 0, &  \hat{\vec{J}}_{\alpha}^{\phantom{\alpha}0} & = -z^0 \vec{m}_\alpha, & \hat{\vec{J}}_{X}^{\phantom{X}0} & = z^0 \vec{A}_X,\\
	\hat{\vec{J}}_{0}^{\phantom{0}\beta} & = \frac{1}{z^0} \vec{k}^\beta, & \hat{\vec{J}}_{\alpha}^{\phantom{\alpha}\beta} & = \vec{k}^\beta \times \vec{m}_\alpha, & \hat{\vec{J}}_{X}^{\phantom{X}\beta} & = \vec{A}_X \times \vec{k}^\beta + \vec{J}_{X}^{\phantom{X}Z}(\vec{A}_Z \cdot \vec{k}^\beta),\\
	\hat{\vec{J}}_{0}^{\phantom{0}Y} & = 0, &  \hat{\vec{J}}_{\alpha}^{\phantom{\alpha}Y} & =0, & \hat{\vec{J}}_{X}^{\phantom{X}Y} & = \vec{J}_{X}^{\phantom{X}Y}.
\end{align}
The last equation means that the components $X,Y$ of the quaternionic structure form a (almost) quaternionic structure in the small space. The integrability of the hypercomplex structure $\hat{\vec{J}}$ leads to non-trivial conditions on $\vec{k}^\alpha$, $\vec{A}_X$ and $\vec{J}_X^{\phantom{X}Y}$. In the following, we only consider the most important ones; the complete list can be found in \cite{Bergshoeff:2004nf}. First, $\vec{k}^\alpha$ and $\vec{m}_\beta$ are independent of $z^0$ and ``satisfy" the $SU(2)$ algebra, in the sense that:
\begin{equation}
	\label{eq:kmSU(2)}
	\vec{k}^\gamma \times \partial_\gamma \vec{k}^\alpha = \vec{k}^\alpha, \qquad \partial_{[\alpha}\vec{m}_{\beta]} = -\frac12 \vec{m}_\alpha \times \vec{m}_\beta.
\end{equation}
The geometric meaning of these equations will become apparent when we explicitly construct these vectors in the next section. Furthermore, we have:
\begin{align}
	\label{eq:zcovconst1}
	\partial_0 \vec{A}_X & = 0, & & (\partial_\alpha + \vec{m}_\alpha \times) \vec{A}_X + \partial_X \vec{m}_\alpha = 0,\\
	\label{eq:zcovconst2}	
	\partial_0 \vec{J}_X^{\phantom{X}Y} & = 0, & & (\partial_\alpha + \vec{m}_\alpha \times) \vec{J}_X^{\phantom{X}Y} = 0.
\end{align}
Finally, we have:
\begin{equation}
	\partial_{[X} \vec{A}_{Y]} - \frac12 \vec{A}_X \times \vec{A}_Y = -\frac12 h_{Z[X}\vec{J}_{Y]}^{\phantom{Y]}Z},
\end{equation}
where $h_{XY}$ is the induced metric\footnote{In the equation that follows, we are assuming that there is a metric $\hat{g}_{XY}$ on the big space. However, $h_{XY}$ can be defined independently of whether a good metric exists or not, we refer again to \cite{Bergshoeff:2004nf} for further details.} on the small space defined as:
\begin{equation}
	h_{XY} = \frac{1}{z^0} \hat{g}_{XY} + \vec{A}_X \cdot \vec{A}_Y.
\end{equation}

It can be shown that the remaining conditions imply that the small space is quaternionic, and that its spin connection $\vec{\omega}_X$ can be chosen to be:
\begin{equation}
	\vec{\omega}_X = -\frac12 \vec{A}_X.
\end{equation}
\subsection{The map from quaternionic to hypercomplex}
\label{sec:embedding}
In the previous section, we have shown that a hypercomplex manifold with conformal symmetry can be related to a ``small" quaternionic manifold. We now construct the inverse map, that is, an explicit prescription to embed a quaternionic manifold characterized by the $SU(2)$ spin connection $\vec{\omega}_X$ and complex structures $\vec{J}_X^{\phantom{X}Y}$ into a hypercomplex manifold with conformal symmetry. This construction is outlined in \cite{Bergshoeff:2004nf}, here we will construct the embedding explicitly. To conform with the notation of the previous section, the extra coordinates are labeled by $z^0$ and $z^\alpha$, so that $q^{\hat X} = (z^0,z^\alpha,q^X)$, with the $q^X$ being the coordinates on the small space.
The $\vec{k}^\alpha(z^\alpha,q^X)$ need to be left-invariant vector fields on $SU(2)$. The dependence on $q^X$ is not fixed at this point, but in the following we will take these vectors to be independent of $q^X$. This means that our construction differs slightly from the one outlined in \cite{Bergshoeff:2004nf}. The advantage will be that we can construct the hypercomplex manifold explicitly given the quaternionic data. We first need to introduce an explicit parametrization of $SU(2)$. We use the Euler parametrization:
\begin{equation}
	U(\psi,\theta,\varphi) = e^{i \sigma^3 \varphi} e^{i \sigma^2 \theta} e^{i \sigma^3 \psi} =
	\begin{pmatrix}
		\cos\frac{\theta}2\,e^{\frac{i}2 (\psi+\varphi)} & \sin\frac{\theta}2\,e^{-\frac{i}2 (\psi-\varphi)} \\[5pt] -\sin\frac{\theta}2\,e^{\frac{i}2 (\psi-\varphi)} & \cos\frac{\theta}2\,e^{-\frac{i}2 (\psi+\varphi)}
	\end{pmatrix},
\end{equation}
so that the (left-invariant) Maurer-Cartan forms $(L^1,L^2,L^3)$ and left-invariant vector fields $(\xi^1,\xi^2,\xi^3)$ read:
\begin{align}
\label{eq:leftinvforms}
\begin{split}
	L^1 &= \sin \varphi\, d\theta - \cos\varphi\,\sin\theta\,d\psi,\\
	L^2 &= \cos\varphi\,d\theta + \sin\varphi\,\sin\theta\,d\psi,\\
	L^3 &= d\varphi + \cos\theta\,d\psi,
\end{split}
\end{align}
and:
\begin{align}
\label{eq:leftinvvect}
\begin{split}
	\xi^1 &= -\frac{\cos\varphi}{\sin{\theta}}\frac{\partial}{\partial\psi} + \sin\varphi \frac{\partial}{\partial\theta} + \cot\theta\,\cos\varphi\frac{\partial}{\partial\varphi},\\
	\xi^2 &= \frac{\sin\varphi}{\sin\theta}\frac{\partial}{\partial\psi} + \cos\varphi\frac{\partial}{\partial\theta}-\cot\theta\,\sin\varphi\frac{\partial}{\partial\varphi},\\
	\xi^3 &= \frac{\partial}{\partial\varphi}.
\end{split}
\end{align}
In particular, notice the important property (which follows from the definition of the Maurer-Cartan forms):
\begin{equation}
	\label{eq:leftMCdef}
	dU = \frac{i}2 L^k \sigma^k U.
\end{equation}
We set:
\begin{equation}
	\vec{k}^\alpha\partial_\alpha = \vec{\xi},
\end{equation}
where the $\vec{\xi}=(\xi^1,\xi^2,\xi^3)$ is the triplet of vectors defined in \eqref{eq:leftinvvect} and we have identified $(z^1,z^2,z^3) \equiv (\psi,\theta,\phi)$. Analogously, the $\vec{m}_\alpha$ are defined from the $L$'s in \eqref{eq:leftinvforms} as:
\begin{equation}
	\vec{m}_\alpha dz^\alpha = \vec{L}.
\end{equation}
It is now evident that $\vec{k}$ and $\vec{m}$ satisfy the relations \eqref{eq:kmSU(2)}, since the latter reduce to the conditions:
\begin{equation}
	[\xi^r,\xi^s]=\epsilon^{rst}\xi^t, \qquad dL^r = -\frac12 \epsilon^{rst} L^s \wedge L^t,
\end{equation}
which are nothing else than the definitions of left-invariant vector fields and left-invariant 1-forms respectively.

Recall that the vector $\vec{A}_X \equiv \frac{1}{z^0} \hat{\vec{J}}_X^{\phantom{X}0}$ induces the spin connection on the small space after gauge-fixing, which is accomplished by taking the $z^\alpha$ to be constants $z^\alpha_0$. Therefore we need to set:
\begin{equation}
	\vec{A}_X(z^\alpha_0,q^X) = -2 \vec{\omega}_X(q^X).
\end{equation}
Analogously, we take:
\begin{equation}
	\hat{\vec{J}}_X^{\phantom{Z}Y}(z^\alpha_0,q^X) = \vec{J}_X^{\phantom{X}Y}(q^X).
\end{equation}
The $z^\alpha$ dependence of these quantities is essentially fixed by the requirement that the complex structures of the big space are integrable. In particular we need to satisfy \eqref{eq:zcovconst1}-\eqref{eq:zcovconst2}.
We therefore take $\vec{A}_X$ and $\vec{J}_X^{\phantom{X}Y}$ to be independent of $z^0$. 
We then need to know how $SU(2)$ transformations act on $SO(3)$ indices. We have:
\begin{equation}
	\label{eq:su2toso3}
	{A_i}^j = i {\vec{\sigma}_i}^{\;j} \cdot \vec{A}, \qquad \vec{A} = -\frac12 i {\vec{\sigma}_i}^{\;j} {A_i}^j.
\end{equation}
As a consequence, given a $SU(2)$ transformation $U$, the corresponding $SO(3)$ transformation $M^{rs}$ is given by:
\begin{equation}
	\label{eq:Mdef}
	M^{rs} = \frac12 \mathrm{Tr}(U\, \sigma^r\, U^\dagger\, \sigma^s),
\end{equation}
where the trace is taken over the $SU(2)$ indices. Using \eqref{eq:leftMCdef}, we easily see that:
\begin{equation}
	dM^{rs} = -\frac12\epsilon^{skt} L^k M^{rt}.
\end{equation}
This means that we can always ``covariantize'' a fixed (that is, independent of the Euler angles) $SO(3)$ vector $\vec{A}$ by considering $\vec{A}_{cov} = M^T \cdot \vec{A} \equiv M^{rs}A^r$. In fact it is easy to check that:
\begin{equation}
	\label{eq:Mcov}
	(d + \vec{L} \times)\vec{A}_{cov} = 0,
\end{equation}
where we have defined the vector of 1-forms $\vec{L} = (L_1,L_2,L_3)$.

In view of property \eqref{eq:Mcov}, we can use the matrix $M$ defined in \eqref{eq:Mdef} to extend $\vec{A}_X$ and $\vec{J}_X^{\phantom{X}Y}$ to general $z$:
\begin{align}
	\vec{A}_X(z^\alpha,q^X) & = (M^T(z^\alpha) M_0)\cdot \vec{A}_X(z^\alpha_0,q^X),\\
	\vec{J}_X^{\phantom{X}Y}(z^\alpha,q^X) & = (M^T(z^\alpha) M_0)\cdot \vec{J}_X^{\phantom{X}Y}(q^X),
\end{align}
where $M_0$ is $M(z^\alpha_0)$, so that when we gauge fix we obtain the spin connection and complex structures we started with. With this choice, $\vec{k}^\alpha$ does not depend on $q^X$. This is basically all we need, and the big space is hypercomplex \cite{Bergshoeff:2004nf}.\footnote{Once again, there are subtleties when the manifolds do not have a metric, but these are not important for our purposes and we refer to \cite{Bergshoeff:2004nf} for a detailed discussion.} Notice that while we always refer to the Euler parametrization, the definitions above are valid in general. However, the analysis above shows that once we gauge fix the extra coordinates associated to the compensator hypermultiplet, we can always choose a particularly convenient gauge as follows:
\begin{equation}
z^1_0 = z^3_0 = 0, \qquad z^2_0 = \frac32 \pi,
\end{equation}
which leads to the conditions:
\begin{equation}
	k^{r\alpha} = \delta^{r\alpha}, \qquad m^{r\alpha} = \delta^{r\alpha}.
\end{equation}
This can be accomplished for \emph{arbitrary} small quaternionic spaces, justifying the choice made in \eqref{eq:gaugechoice}.
\subsection{Metric and symmetries}\label{sec:apphypersymmetries}
When we have a metric on the small space (so that it is quaternionic-K\"ahler), we can also construct a metric on the big space as follows. We use $\hat A = (i,A)$, where $i=1,2$. First, the vielbein on the large space is determined from \eqref{eq:hattedquaternionicstructure} to be:
\begin{equation}
  \begin{array}{lll}
\widehat f^0_{ij} = -\rmi\varepsilon _{ij}\sqrt{\ft12z^0}, \qquad &
  \widehat f^\alpha_{ij} =\sqrt{\frac{1}{2z^0}}\vec k^\alpha \cdot \vec
\sigma_{ij},\qquad &
 \widehat f^X_{ij}=0, \\
\widehat f^0_{iA}=0,\qquad & \widehat{f}^\alpha _{iA}=f^X_{iA}\vec A_X\cdot \vec k^\alpha,\qquad & \widehat f_{iA}^X=f_{iA}^X,\\
 \widehat f_0^{ij}=\rmi\varepsilon^{ij}\sqrt{\frac{1}{2z^0}},\qquad & \widehat f_\alpha^{ij}
  = - \sqrt{\frac{z^0}{2}} \vm\alpha\cdot
\vec\sigma^{ij}
 ,\qquad &\widehat f_X^{ij}= \sqrt{\frac{z^0}{2}} \vec A_X\cdot
\vec\sigma^{ij} ,
 \\
\widehat f_0^{iA}=0,\qquad  & \widehat f_\alpha^{iA}=0,\qquad &\widehat
f_X^{iA}=f_X^{iA}.
  \end{array} \label{allf}
\end{equation}
Notice that the index structure in the definition of Pauli matrices is taken to be $\vec{\sigma}_i^{\phantom{i}j}$, so the indices in the Pauli matrices used above are raised and lowered using the $\epsilon$ tensor with the usual conventions. As explained in the first section, the metric can be obtained by defining an appropriate matrix $\hat{C}_{\hat{A}\hat{B}}$. We take:
\begin{equation}
	\hat{C}_{AB}=C_{AB},\qquad \hat{C}_{ij} = \epsilon_{ij}, \qquad \hat{C}_{iA} = 0.
\end{equation}
The metric on the big space is then simply:
\begin{align}
\begin{split}
	d\hat{s}^2 = - \frac{(dz^0)^2}{z^0} +  \Big\{ & z^0 h_{XY}(q) dq^X dq^Y  \\ &  + \hat{g}_{\alpha\beta} \left(dz^\alpha - \vec{A}_X(z,q) \cdot \vec{k}^\alpha dq^X \right) \left(dz^\beta - \vec{A}_Y(z,q) \cdot \vec{k}^\beta dq^Y \right) \Big\},
\end{split}
\end{align}
where $g_{XY} = z^0 h_{XY}$ is the metric induced on the small space.

Now we briefly turn to symmetries, which are important when we consider gauged models. Working once again with spaces equipped with a metric, symmetries can be characterized by vectors $k^X_I$:
\begin{equation}
	\delta q^X = \Lambda^I k^X_I(q),
\end{equation}
which satisfy the Killing equation:
\begin{equation}
	\label{eq:isokilling}
	\mathfrak{D}_{(X} k_{Y)I}  =0.
\end{equation}
The moment maps $\nu \vec{P}_I(q)$ are defined via:
\begin{equation}
	\nu \vec{P}_I = - \frac{1}{4r} \vec{J}_X^{\phantom{X}Y} \mathfrak{D}_Y k^X_I.
\end{equation}
The extension of these symmetries to the hyperk\"ahler manifold is straightforward. Let us define $\hat{k}^{\hat X}_I$ as:
\begin{equation}
	\label{eq:liftedk}
	\hat{k}^0_I = 0, \qquad \hat{k}^\alpha_I = \vec{k}^\alpha \cdot \vec{r}_I, \qquad \hat{k}^X_I = k^X_I,	
\end{equation}
where $\vec{r}_I$ is defined by:
\begin{equation}
	\label{eq:normalizingsymmetry}
	\mathcal{L}_{k_I} \vec{J}_X^{\phantom{X}Y} = \vec{r}_I \times \vec{J}_X^{\phantom{X}Y}.
\end{equation}
Equivalently, it is not difficult to show that:
\begin{equation}
	\label{eq:vecrI}
	\vec{r}_I = - \frac{1}{8r} \vec{J}_Y^{\phantom{Y}X} \times \mathcal{L}_{k_I} \vec{J}_X^{\phantom{X}Y}.
\end{equation}
As a consequence, we have:
\begin{equation}
	\label{eq:prepot}
	\nu \vec{P}_I = -\frac12 \vec{r}_I - k_I^X \vec{\omega}_X.
\end{equation}

\vspace*{0.2in}

\bibliographystyle{utphys}
\bibliography{hdcmax}

\providecommand{\href}[2]{#2}\begingroup\raggedright\begin{thebibliography}{10}

\bibitem{Henningson:1998gx}
M.~Henningson and K.~Skenderis, ``{The Holographic Weyl anomaly},'' {\em JHEP}
  {\bf 9807} (1998)  023,
\href{http://arxiv.org/abs/hep-th/9806087}{{\tt arXiv:hep-th/9806087
  [hep-th]}}.

\bibitem{Aharony:1999rz}
O.~Aharony, J.~Pawelczyk, S.~Theisen, and S.~Yankielowicz, ``{A Note on
  anomalies in the AdS / CFT correspondence},''
  \href{http://dx.doi.org/10.1103/PhysRevD.60.066001}{{\em Phys.Rev.} {\bf D60}
  (1999)  066001},
\href{http://arxiv.org/abs/hep-th/9901134}{{\tt arXiv:hep-th/9901134
  [hep-th]}}.

\bibitem{Anselmi:1998zb}
D.~Anselmi and A.~Kehagias, ``{Subleading corrections and central charges in
  the AdS / CFT correspondence},''
  \href{http://dx.doi.org/10.1016/S0370-2693(99)00446-3}{{\em Phys.Lett.} {\bf
  B455} (1999)  155--163},
\href{http://arxiv.org/abs/hep-th/9812092}{{\tt arXiv:hep-th/9812092
  [hep-th]}}.

\bibitem{Bilal:1999ty}
A.~Bilal and C.-S. Chu, ``{Testing the AdS / CFT correspondence beyond large
  N},'' {\em PoS} {\bf tmr99} (1999)  009,
\href{http://arxiv.org/abs/hep-th/0003129}{{\tt arXiv:hep-th/0003129
  [hep-th]}}.

\bibitem{Bilal:1999ph}
A.~Bilal and C.-S. Chu, ``{A Note on the chiral anomaly in the AdS / CFT
  correspondence and 1 / N**2 correction},''
  \href{http://dx.doi.org/10.1016/S0550-3213(99)00553-2}{{\em Nucl.Phys.} {\bf
  B562} (1999)  181--190},
\href{http://arxiv.org/abs/hep-th/9907106}{{\tt arXiv:hep-th/9907106
  [hep-th]}}.

\bibitem{Mansfield:2000zw}
P.~Mansfield and D.~Nolland, ``{Order 1 / N**2 test of the Maldacena
  conjecture: Cancellation of the one loop Weyl anomaly},''
  \href{http://dx.doi.org/10.1016/S0370-2693(00)01247-8}{{\em Phys.Lett.} {\bf
  B495} (2000)  435--439},
\href{http://arxiv.org/abs/hep-th/0005224}{{\tt arXiv:hep-th/0005224
  [hep-th]}}.

\bibitem{Mansfield:2002pa}
P.~Mansfield, D.~Nolland, and T.~Ueno, ``{Order 1 / N**2 test of the Maldacena
  conjecture. 2. The Full bulk one loop contribution to the boundary Weyl
  anomaly},'' \href{http://dx.doi.org/10.1016/S0370-2693(03)00750-0}{{\em
  Phys.Lett.} {\bf B565} (2003)  207--210},
\href{http://arxiv.org/abs/hep-th/0208135}{{\tt arXiv:hep-th/0208135
  [hep-th]}}.

\bibitem{Mansfield:2003gs}
P.~Mansfield, D.~Nolland, and T.~Ueno, ``{The Boundary Weyl anomaly in the N=4
  SYM / type IIB supergravity correspondence},''
  \href{http://dx.doi.org/10.1088/1126-6708/2004/01/013}{{\em JHEP} {\bf 0401}
  (2004)  013},
\href{http://arxiv.org/abs/hep-th/0311021}{{\tt arXiv:hep-th/0311021
  [hep-th]}}.

\bibitem{Liu:2010gz}
J.~T. Liu and R.~Minasian, ``{Computing $1/N^2$ corrections in AdS/CFT},''
\href{http://arxiv.org/abs/1010.6074}{{\tt arXiv:1010.6074 [hep-th]}}.

\bibitem{Ardehali:2013gra}
A.~Arabi~Ardehali, J.~T. Liu, and P.~Szepietowski, ``{The spectrum of IIB
  supergravity on $AdS_5$ x $S^5/Z_3$ and a $1/N^2$ test of AdS/CFT},''
  \href{http://dx.doi.org/10.1007/JHEP06(2013)024}{{\em JHEP} {\bf 1306} (2013)
   024},
\href{http://arxiv.org/abs/1304.1540}{{\tt arXiv:1304.1540 [hep-th]}}.

\bibitem{Ardehali:2013xya}
A.~A. Ardehali, J.~T. Liu, and P.~Szepietowski, ``{$1/N^2$ corrections to the
  holographic Weyl anomaly},''
  \href{http://dx.doi.org/10.1007/JHEP01(2014)002}{{\em JHEP} {\bf 1401} (2014)
   002},
\href{http://arxiv.org/abs/1310.2611}{{\tt arXiv:1310.2611 [hep-th]}}.

\bibitem{Ardehali:2013xla}
A.~Arabi~Ardehali, J.~T. Liu, and P.~Szepietowski, ``{The shortened KK spectrum
  of IIB supergravity on $Y^{p,q}$},''
  \href{http://dx.doi.org/10.1007/JHEP02(2014)064}{{\em JHEP} {\bf 1402} (2014)
   064},
\href{http://arxiv.org/abs/1311.4550}{{\tt arXiv:1311.4550 [hep-th]}}.

\bibitem{Tseytlin:2000sf}
A.~A. Tseytlin, ``{R**4 terms in 11 dimensions and conformal anomaly of (2,0)
  theory},'' \href{http://dx.doi.org/10.1016/S0550-3213(00)00380-1}{{\em
  Nucl.Phys.} {\bf B584} (2000)  233--250},
\href{http://arxiv.org/abs/hep-th/0005072}{{\tt arXiv:hep-th/0005072
  [hep-th]}}.

\bibitem{Intriligator:2003jj}
K.~A. Intriligator and B.~Wecht, ``{The Exact superconformal R symmetry
  maximizes a},'' \href{http://dx.doi.org/10.1016/S0550-3213(03)00459-0}{{\em
  Nucl.Phys.} {\bf B667} (2003)  183--200},
\href{http://arxiv.org/abs/hep-th/0304128}{{\tt arXiv:hep-th/0304128
  [hep-th]}}.

\bibitem{Benini:2012cz}
F.~Benini and N.~Bobev, ``{Exact two-dimensional superconformal R-symmetry and
  c-extremization},''
  \href{http://dx.doi.org/10.1103/PhysRevLett.110.061601}{{\em Phys.Rev.Lett.}
  {\bf 110} (2013) no.~6, 061601},
\href{http://arxiv.org/abs/1211.4030}{{\tt arXiv:1211.4030 [hep-th]}}.

\bibitem{Tachikawa:2005tq}
Y.~Tachikawa, ``{Five-dimensional supergravity dual of a-maximization},''
  \href{http://dx.doi.org/10.1016/j.nuclphysb.2005.11.010}{{\em Nucl.Phys.}
  {\bf B733} (2006)  188--203},
\href{http://arxiv.org/abs/hep-th/0507057}{{\tt arXiv:hep-th/0507057
  [hep-th]}}.

\bibitem{Hanaki:2006pj}
K.~Hanaki, K.~Ohashi, and Y.~Tachikawa, ``{Supersymmetric Completion of an R**2
  term in Five-dimensional Supergravity},''
  \href{http://dx.doi.org/10.1143/PTP.117.533}{{\em Prog.Theor.Phys.} {\bf 117}
  (2007)  533},
\href{http://arxiv.org/abs/hep-th/0611329}{{\tt arXiv:hep-th/0611329
  [hep-th]}}.

\bibitem{Szepietowski:2012tb}
P.~Szepietowski, ``{Comments on a-maximization from gauged supergravity},''
  \href{http://dx.doi.org/10.1007/JHEP12(2012)018}{{\em JHEP} {\bf 1212} (2012)
   018},
\href{http://arxiv.org/abs/1209.3025}{{\tt arXiv:1209.3025 [hep-th]}}.

\bibitem{Karndumri:2013iqa}
P.~Karndumri and E.~O~Colgain, ``{Supergravity dual of $c$-extremization},''
  \href{http://dx.doi.org/10.1103/PhysRevD.87.101902}{{\em Phys.Rev.} {\bf D87}
  (2013) no.~10, 101902},
\href{http://arxiv.org/abs/1302.6532}{{\tt arXiv:1302.6532 [hep-th]}}.

\bibitem{Gaiotto:2009we}
D.~Gaiotto, ``{N=2 dualities},''
  \href{http://dx.doi.org/10.1007/JHEP08(2012)034}{{\em JHEP} {\bf 1208} (2012)
   034},
\href{http://arxiv.org/abs/0904.2715}{{\tt arXiv:0904.2715 [hep-th]}}.

\bibitem{Gaiotto:2009hg}
D.~Gaiotto, G.~W. Moore, and A.~Neitzke, ``{Wall-crossing, Hitchin Systems, and
  the WKB Approximation},''
\href{http://arxiv.org/abs/0907.3987}{{\tt arXiv:0907.3987 [hep-th]}}.

\bibitem{Alday:2009qq}
L.~F. Alday, F.~Benini, and Y.~Tachikawa, ``{Liouville/Toda central charges
  from M5-branes},''
  \href{http://dx.doi.org/10.1103/PhysRevLett.105.141601}{{\em Phys.Rev.Lett.}
  {\bf 105} (2010)  141601},
\href{http://arxiv.org/abs/0909.4776}{{\tt arXiv:0909.4776 [hep-th]}}.

\bibitem{Kanno:2009ga}
S.~Kanno, Y.~Matsuo, S.~Shiba, and Y.~Tachikawa, ``{N=2 gauge theories and
  degenerate fields of Toda theory},''
  \href{http://dx.doi.org/10.1103/PhysRevD.81.046004}{{\em Phys.Rev.} {\bf D81}
  (2010)  046004},
\href{http://arxiv.org/abs/0911.4787}{{\tt arXiv:0911.4787 [hep-th]}}.

\bibitem{Gaiotto:2010be}
D.~Gaiotto, G.~W. Moore, and A.~Neitzke, ``{Framed BPS States},''
\href{http://arxiv.org/abs/1006.0146}{{\tt arXiv:1006.0146 [hep-th]}}.

\bibitem{Tachikawa:2010vg}
Y.~Tachikawa, ``{N=2 S-duality via Outer-automorphism Twists},''
  \href{http://dx.doi.org/10.1088/1751-8113/44/18/182001}{{\em J.Phys.A} {\bf
  A44} (2011)  182001},
\href{http://arxiv.org/abs/1009.0339}{{\tt arXiv:1009.0339 [hep-th]}}.

\bibitem{Nanopoulos:2010ga}
D.~Nanopoulos and D.~Xie, ``{$N=2$ Generalized Superconformal Quiver Gauge
  Theory},''
\href{http://arxiv.org/abs/1006.3486}{{\tt arXiv:1006.3486 [hep-th]}}.

\bibitem{Chacaltana:2010ks}
O.~Chacaltana and J.~Distler, ``{Tinkertoys for Gaiotto Duality},''
  \href{http://dx.doi.org/10.1007/JHEP11(2010)099}{{\em JHEP} {\bf 1011} (2010)
   099},
\href{http://arxiv.org/abs/1008.5203}{{\tt arXiv:1008.5203 [hep-th]}}.

\bibitem{Gaiotto:2011tf}
D.~Gaiotto, G.~W. Moore, and A.~Neitzke, ``{Wall-Crossing in Coupled 2d-4d
  Systems},''
\href{http://arxiv.org/abs/1103.2598}{{\tt arXiv:1103.2598 [hep-th]}}.

\bibitem{Cecotti:2011rv}
S.~Cecotti and C.~Vafa, ``{Classification of complete N=2 supersymmetric
  theories in 4 dimensions},''
\href{http://arxiv.org/abs/1103.5832}{{\tt arXiv:1103.5832 [hep-th]}}.

\bibitem{Tachikawa:2011yr}
Y.~Tachikawa and S.~Terashima, ``{Seiberg-Witten Geometries Revisited},''
  \href{http://dx.doi.org/10.1007/JHEP09(2011)010}{{\em JHEP} {\bf 1109} (2011)
   010},
\href{http://arxiv.org/abs/1108.2315}{{\tt arXiv:1108.2315 [hep-th]}}.

\bibitem{Alim:2011ae}
M.~Alim, S.~Cecotti, C.~Cordova, S.~Espahbodi, A.~Rastogi, {\em et al.}, ``{BPS
  Quivers and Spectra of Complete N=2 Quantum Field Theories},''
\href{http://arxiv.org/abs/1109.4941}{{\tt arXiv:1109.4941 [hep-th]}}.

\bibitem{Gaiotto:2011xs}
D.~Gaiotto, G.~W. Moore, and Y.~Tachikawa, ``{On 6d N=(2,0) theory compactified
  on a Riemann surface with finite area},''
\href{http://arxiv.org/abs/1110.2657}{{\tt arXiv:1110.2657 [hep-th]}}.

\bibitem{Alim:2011kw}
M.~Alim, S.~Cecotti, C.~Cordova, S.~Espahbodi, A.~Rastogi, {\em et al.}, ``{N=2
  Quantum Field Theories and Their BPS Quivers},''
  \href{http://arxiv.org/abs/1112.3984}{{\tt arXiv:1112.3984 [hep-th]}}.
93 Pages, 18 Figures.

\bibitem{Bah:2011vv}
I.~Bah, C.~Beem, N.~Bobev, and B.~Wecht, ``{AdS/CFT Dual Pairs from M5-Branes
  on Riemann Surfaces},''
  \href{http://dx.doi.org/10.1103/PhysRevD.85.121901}{{\em Phys.Rev.} {\bf D85}
  (2012)  121901},
\href{http://arxiv.org/abs/1112.5487}{{\tt arXiv:1112.5487 [hep-th]}}.

\bibitem{Bah:2012dg}
I.~Bah, C.~Beem, N.~Bobev, and B.~Wecht, ``{Four-Dimensional SCFTs from
  M5-Branes},'' \href{http://dx.doi.org/10.1007/JHEP06(2012)005}{{\em JHEP}
  {\bf 1206} (2012)  005},
\href{http://arxiv.org/abs/1203.0303}{{\tt arXiv:1203.0303 [hep-th]}}.

\bibitem{Bah:2011je}
I.~Bah and B.~Wecht, ``{New N=1 Superconformal Field Theories In Four
  Dimensions},'' \href{http://dx.doi.org/10.1007/JHEP07(2013)107}{{\em JHEP}
  {\bf 1307} (2013)  107},
\href{http://arxiv.org/abs/1111.3402}{{\tt arXiv:1111.3402 [hep-th]}}.

\bibitem{Benini:2013cda}
F.~Benini and N.~Bobev, ``{Two-dimensional SCFTs from wrapped branes and
  c-extremization},'' \href{http://dx.doi.org/10.1007/JHEP06(2013)005}{{\em
  JHEP} {\bf 1306} (2013)  005},
\href{http://arxiv.org/abs/1302.4451}{{\tt arXiv:1302.4451 [hep-th]}}.

\bibitem{Maldacena:2000mw}
J.~M. Maldacena and C.~Nunez, ``{Supergravity description of field theories on
  curved manifolds and a no go theorem},''
  \href{http://dx.doi.org/10.1142/S0217751X01003937}{{\em Int.J.Mod.Phys.} {\bf
  A16} (2001)  822--855},
\href{http://arxiv.org/abs/hep-th/0007018}{{\tt arXiv:hep-th/0007018
  [hep-th]}}.

\bibitem{Gauntlett:2001jj}
J.~P. Gauntlett and N.~Kim, ``{M five-branes wrapped on supersymmetric cycles.
  2.},'' \href{http://dx.doi.org/10.1103/PhysRevD.65.086003}{{\em Phys.Rev.}
  {\bf D65} (2002)  086003},
\href{http://arxiv.org/abs/hep-th/0109039}{{\tt arXiv:hep-th/0109039
  [hep-th]}}.

\bibitem{Castro:2007sd}
A.~Castro, J.~L. Davis, P.~Kraus, and F.~Larsen, ``{5D attractors with higher
  derivatives},'' \href{http://dx.doi.org/10.1088/1126-6708/2007/04/091}{{\em
  JHEP} {\bf 0704} (2007)  091},
\href{http://arxiv.org/abs/hep-th/0702072}{{\tt arXiv:hep-th/0702072
  [hep-th]}}.

\bibitem{Ozkan:2013nwa}
M.~Ozkan and Y.~Pang, ``{All off-shell $R^{2}$ invariants in five dimensional
  $\mathcal{N} =$ 2 supergravity},''
  \href{http://dx.doi.org/10.1007/JHEP08(2013)042}{{\em JHEP} {\bf 1308} (2013)
   042},
\href{http://arxiv.org/abs/1306.1540}{{\tt arXiv:1306.1540}}.

\bibitem{Strominger:1997eb}
A.~Strominger, ``{Loop corrections to the universal hypermultiplet},''
  \href{http://dx.doi.org/10.1016/S0370-2693(98)00015-X}{{\em Phys.Lett.} {\bf
  B421} (1998)  139--148},
\href{http://arxiv.org/abs/hep-th/9706195}{{\tt arXiv:hep-th/9706195
  [hep-th]}}.

\bibitem{Gunther:1998sc}
H.~Gunther, C.~Herrmann, and J.~Louis, ``{Quantum corrections in the
  hypermultiplet moduli space},'' {\em Fortsch.Phys.} {\bf 48} (2000)
  119--123,
\href{http://arxiv.org/abs/hep-th/9901137}{{\tt arXiv:hep-th/9901137
  [hep-th]}}.

\bibitem{Antoniadis:2003sw}
I.~Antoniadis, R.~Minasian, S.~Theisen, and P.~Vanhove, ``{String loop
  corrections to the universal hypermultiplet},''
  \href{http://dx.doi.org/10.1088/0264-9381/20/23/009}{{\em Class.Quant.Grav.}
  {\bf 20} (2003)  5079--5102},
\href{http://arxiv.org/abs/hep-th/0307268}{{\tt arXiv:hep-th/0307268
  [hep-th]}}.

\bibitem{RoblesLlana:2006ez}
D.~Robles-Llana, F.~Saueressig, and S.~Vandoren, ``{String loop corrected
  hypermultiplet moduli spaces},''
  \href{http://dx.doi.org/10.1088/1126-6708/2006/03/081}{{\em JHEP} {\bf 0603}
  (2006)  081},
\href{http://arxiv.org/abs/hep-th/0602164}{{\tt arXiv:hep-th/0602164
  [hep-th]}}.

\bibitem{Alexandrov:2007ec}
S.~Alexandrov, ``{Quantum covariant c-map},''
  \href{http://dx.doi.org/10.1088/1126-6708/2007/05/094}{{\em JHEP} {\bf 0705}
  (2007)  094},
\href{http://arxiv.org/abs/hep-th/0702203}{{\tt arXiv:hep-th/0702203
  [HEP-TH]}}.

\bibitem{Harvey:1998bx}
J.~A. Harvey, R.~Minasian, and G.~W. Moore, ``{NonAbelian tensor multiplet
  anomalies},'' {\em JHEP} {\bf 9809} (1998)  004,
\href{http://arxiv.org/abs/hep-th/9808060}{{\tt arXiv:hep-th/9808060
  [hep-th]}}.

\bibitem{Yi:2001bz}
P.~Yi, ``{Anomaly of (2,0) theories},''
  \href{http://dx.doi.org/10.1103/PhysRevD.64.106006}{{\em Phys.Rev.} {\bf D64}
  (2001)  106006},
\href{http://arxiv.org/abs/hep-th/0106165}{{\tt arXiv:hep-th/0106165
  [hep-th]}}.

\bibitem{Intriligator:2000eq}
K.~A. Intriligator, ``{Anomaly matching and a Hopf-Wess-Zumino term in 6d,
  N=(2,0) field theories},''
  \href{http://dx.doi.org/10.1016/S0550-3213(00)00148-6}{{\em Nucl.Phys.} {\bf
  B581} (2000)  257--273},
\href{http://arxiv.org/abs/hep-th/0001205}{{\tt arXiv:hep-th/0001205
  [hep-th]}}.

\bibitem{deWit:1984px}
B.~de~Wit, P.~Lauwers, and A.~Van~Proeyen, ``{Lagrangians of $N=2$ Supergravity
  - Matter Systems},''
\href{http://dx.doi.org/10.1016/0550-3213(85)90154-3}{{\em Nucl.Phys.} {\bf
  B255} (1985)  569}.

\bibitem{Bergshoeff:2004kh}
E.~Bergshoeff, S.~Cucu, T.~de~Wit, J.~Gheerardyn, S.~Vandoren, {\em et al.},
  ``{N = 2 supergravity in five-dimensions revisited},''
  \href{http://dx.doi.org/10.1088/0264-9381/23/23/C01}{{\em Class.Quant.Grav.}
  {\bf 21} (2004)  3015--3042},
\href{http://arxiv.org/abs/hep-th/0403045}{{\tt arXiv:hep-th/0403045
  [hep-th]}}.

\bibitem{Fujita:2001kv}
T.~Fujita and K.~Ohashi, ``{Superconformal tensor calculus in
  five-dimensions},'' \href{http://dx.doi.org/10.1143/PTP.106.221}{{\em
  Prog.Theor.Phys.} {\bf 106} (2001)  221--247},
\href{http://arxiv.org/abs/hep-th/0104130}{{\tt arXiv:hep-th/0104130
  [hep-th]}}.

\bibitem{Castro:2007hc}
A.~Castro, J.~L. Davis, P.~Kraus, and F.~Larsen, ``{5D Black Holes and Strings
  with Higher Derivatives},''
  \href{http://dx.doi.org/10.1088/1126-6708/2007/06/007}{{\em JHEP} {\bf 0706}
  (2007)  007},
\href{http://arxiv.org/abs/hep-th/0703087}{{\tt arXiv:hep-th/0703087
  [hep-th]}}.

\bibitem{Behrndt:2001km}
K.~Behrndt and G.~Dall'Agata, ``{Vacua of N=2 gauged supergravity derived from
  nonhomogenous quaternionic spaces},''
  \href{http://dx.doi.org/10.1016/S0550-3213(02)00053-6}{{\em Nucl.Phys.} {\bf
  B627} (2002)  357--380},
\href{http://arxiv.org/abs/hep-th/0112136}{{\tt arXiv:hep-th/0112136
  [hep-th]}}.

\bibitem{Blau:1999vz}
M.~Blau, K.~Narain, and E.~Gava, ``{On subleading contributions to the AdS /
  CFT trace anomaly},''
  \href{http://dx.doi.org/10.1088/1126-6708/1999/09/018}{{\em JHEP} {\bf 9909}
  (1999)  018},
\href{http://arxiv.org/abs/hep-th/9904179}{{\tt arXiv:hep-th/9904179
  [hep-th]}}.

\bibitem{Nojiri:1999mh}
S.~Nojiri and S.~D. Odintsov, ``{On the conformal anomaly from higher
  derivative gravity in AdS / CFT correspondence},''
  \href{http://dx.doi.org/10.1142/S0217751X00000197}{{\em Int.J.Mod.Phys.} {\bf
  A15} (2000)  413--428},
\href{http://arxiv.org/abs/hep-th/9903033}{{\tt arXiv:hep-th/9903033
  [hep-th]}}.

\bibitem{Fukuma:2001uf}
M.~Fukuma, S.~Matsuura, and T.~Sakai, ``{Higher derivative gravity and the AdS
  / CFT correspondence},'' \href{http://dx.doi.org/10.1143/PTP.105.1017}{{\em
  Prog.Theor.Phys.} {\bf 105} (2001)  1017--1044},
\href{http://arxiv.org/abs/hep-th/0103187}{{\tt arXiv:hep-th/0103187
  [hep-th]}}.

\bibitem{Cremonini:2008tw}
S.~Cremonini, K.~Hanaki, J.~T. Liu, and P.~Szepietowski, ``{Black holes in
  five-dimensional gauged supergravity with higher derivatives},''
  \href{http://dx.doi.org/10.1088/1126-6708/2009/12/045}{{\em JHEP} {\bf 0912}
  (2009)  045},
\href{http://arxiv.org/abs/0812.3572}{{\tt arXiv:0812.3572 [hep-th]}}.

\bibitem{Brown:1986nw}
J.~D. Brown and M.~Henneaux, ``{Central Charges in the Canonical Realization of
  Asymptotic Symmetries: An Example from Three-Dimensional Gravity},''
\href{http://dx.doi.org/10.1007/BF01211590}{{\em Commun.Math.Phys.} {\bf 104}
  (1986)  207--226}.

\bibitem{Imbimbo:1999bj}
C.~Imbimbo, A.~Schwimmer, S.~Theisen, and S.~Yankielowicz, ``{Diffeomorphisms
  and holographic anomalies},''
  \href{http://dx.doi.org/10.1088/0264-9381/17/5/322}{{\em Class.Quant.Grav.}
  {\bf 17} (2000)  1129--1138},
\href{http://arxiv.org/abs/hep-th/9910267}{{\tt arXiv:hep-th/9910267
  [hep-th]}}.

\bibitem{Saida:1999ec}
H.~Saida and J.~Soda, ``{Statistical entropy of BTZ black hole in higher
  curvature gravity},''
  \href{http://dx.doi.org/10.1016/S0370-2693(99)01405-7}{{\em Phys.Lett.} {\bf
  B471} (2000)  358--366},
\href{http://arxiv.org/abs/gr-qc/9909061}{{\tt arXiv:gr-qc/9909061 [gr-qc]}}.

\bibitem{Kraus:2005vz}
P.~Kraus and F.~Larsen, ``{Microscopic black hole entropy in theories with
  higher derivatives},''
  \href{http://dx.doi.org/10.1088/1126-6708/2005/09/034}{{\em JHEP} {\bf 0509}
  (2005)  034},
\href{http://arxiv.org/abs/hep-th/0506176}{{\tt arXiv:hep-th/0506176
  [hep-th]}}.

\bibitem{Kraus:2006wn}
P.~Kraus, ``{Lectures on black holes and the AdS(3) / CFT(2) correspondence},''
  {\em Lect.Notes Phys.} {\bf 755} (2008)  193--247,
\href{http://arxiv.org/abs/hep-th/0609074}{{\tt arXiv:hep-th/0609074
  [hep-th]}}.

\bibitem{Kraus:2005zm}
P.~Kraus and F.~Larsen, ``{Holographic gravitational anomalies},''
  \href{http://dx.doi.org/10.1088/1126-6708/2006/01/022}{{\em JHEP} {\bf 0601}
  (2006)  022},
\href{http://arxiv.org/abs/hep-th/0508218}{{\tt arXiv:hep-th/0508218
  [hep-th]}}.

\bibitem{Solodukhin:2005ah}
S.~N. Solodukhin, ``{Holography with gravitational Chern-Simons},''
  \href{http://dx.doi.org/10.1103/PhysRevD.74.024015}{{\em Phys.Rev.} {\bf D74}
  (2006)  024015},
\href{http://arxiv.org/abs/hep-th/0509148}{{\tt arXiv:hep-th/0509148
  [hep-th]}}.

\bibitem{Solodukhin:2005ns}
S.~N. Solodukhin, ``{Holographic description of gravitational anomalies},''
  \href{http://dx.doi.org/10.1088/1126-6708/2006/07/003}{{\em JHEP} {\bf 0607}
  (2006)  003},
\href{http://arxiv.org/abs/hep-th/0512216}{{\tt arXiv:hep-th/0512216
  [hep-th]}}.

\bibitem{Karndumri:2013dca}
P.~Karndumri and E.~O. Colgáin, ``{3D Supergravity from wrapped D3-branes},''
  \href{http://dx.doi.org/10.1007/JHEP10(2013)094}{{\em JHEP} {\bf 1310} (2013)
   094},
\href{http://arxiv.org/abs/1307.2086}{{\tt arXiv:1307.2086}}.

\bibitem{Kinney:2005ej}
J.~Kinney, J.~M. Maldacena, S.~Minwalla, and S.~Raju, ``{An Index for 4
  dimensional super conformal theories},''
  \href{http://dx.doi.org/10.1007/s00220-007-0258-7}{{\em Commun.Math.Phys.}
  {\bf 275} (2007)  209--254},
\href{http://arxiv.org/abs/hep-th/0510251}{{\tt arXiv:hep-th/0510251
  [hep-th]}}.

\bibitem{Chang:2013fba}
C.-M. Chang and X.~Yin, ``{1/16 BPS states in $\mathcal N=$ 4 super-Yang-Mills
  theory},'' \href{http://dx.doi.org/10.1103/PhysRevD.88.106005}{{\em
  Phys.Rev.} {\bf D88} (2013) no.~10, 106005},
\href{http://arxiv.org/abs/1305.6314}{{\tt arXiv:1305.6314 [hep-th]}}.

\bibitem{Bergshoeff:2001hc}
E.~Bergshoeff, S.~Cucu, M.~Derix, T.~de~Wit, R.~Halbersma, {\em et al.},
  ``{Weyl multiplets of N=2 conformal supergravity in five-dimensions},''
  \href{http://dx.doi.org/10.1088/1126-6708/2001/06/051}{{\em JHEP} {\bf 0106}
  (2001)  051},
\href{http://arxiv.org/abs/hep-th/0104113}{{\tt arXiv:hep-th/0104113
  [hep-th]}}.

\bibitem{deWit:2001dj}
B.~de~Wit, M.~Rocek, and S.~Vandoren, ``{Hypermultiplets, hyperKahler cones and
  quaternion Kahler geometry},''
  \href{http://dx.doi.org/10.1088/1126-6708/2001/02/039}{{\em JHEP} {\bf 0102}
  (2001)  039},
\href{http://arxiv.org/abs/hep-th/0101161}{{\tt arXiv:hep-th/0101161
  [hep-th]}}.

\bibitem{Bergshoeff:2002qk}
E.~Bergshoeff, S.~Cucu, T.~De~Wit, J.~Gheerardyn, R.~Halbersma, {\em et al.},
  ``{Superconformal N=2, D = 5 matter with and without actions},''
  \href{http://dx.doi.org/10.1088/1126-6708/2002/10/045}{{\em JHEP} {\bf 0210}
  (2002)  045},
\href{http://arxiv.org/abs/hep-th/0205230}{{\tt arXiv:hep-th/0205230
  [hep-th]}}.

\bibitem{Pernici:1984xx}
M.~Pernici, K.~Pilch, and P.~van Nieuwenhuizen, ``{Guged Maximally Extended
  Supergravity in Seven Dimensions},''
\href{http://dx.doi.org/10.1016/0370-2693(84)90813-X}{{\em Phys.Lett.} {\bf
  B143} (1984)  103}.

\bibitem{Liu:1999ai}
J.~T. Liu and R.~Minasian, ``{Black holes and membranes in AdS(7)},''
  \href{http://dx.doi.org/10.1016/S0370-2693(99)00500-6}{{\em Phys.Lett.} {\bf
  B457} (1999)  39--46},
\href{http://arxiv.org/abs/hep-th/9903269}{{\tt arXiv:hep-th/9903269
  [hep-th]}}.

\bibitem{Cvetic:1999xp}
M.~Cvetic, M.~Duff, P.~Hoxha, J.~T. Liu, H.~Lu, {\em et al.}, ``{Embedding AdS
  black holes in ten-dimensions and eleven-dimensions},''
  \href{http://dx.doi.org/10.1016/S0550-3213(99)00419-8}{{\em Nucl.Phys.} {\bf
  B558} (1999)  96--126},
\href{http://arxiv.org/abs/hep-th/9903214}{{\tt arXiv:hep-th/9903214
  [hep-th]}}.

\bibitem{Bergshoeff:2004nf}
E.~Bergshoeff, S.~Cucu, T.~de~Wit, J.~Gheerardyn, S.~Vandoren, {\em et al.},
  ``{The Map between conformal hypercomplex/hyper-Kahler and
  quaternionic(-Kahler) geometry},''
  \href{http://dx.doi.org/10.1007/s00220-005-1475-6}{{\em Commun.Math.Phys.}
  {\bf 262} (2006)  411--457},
\href{http://arxiv.org/abs/hep-th/0411209}{{\tt arXiv:hep-th/0411209
  [hep-th]}}.

\bibitem{VanProeyen:2001ng}
A.~Van~Proeyen, ``{The Scalars of N=2, D = 5 and attractor equations},''
\href{http://arxiv.org/abs/hep-th/0105158}{{\tt arXiv:hep-th/0105158
  [hep-th]}}.

\end{thebibliography}\endgroup
\end{document}